\documentclass[draftcls, onecolumn, 12pt]{IEEEtran}%

\usepackage{amsmath}
\usepackage{amsthm}
\usepackage{amsfonts}
\usepackage{amssymb}
\usepackage{graphicx}
\usepackage{algorithm}
\usepackage{algpseudocode}  
\usepackage[usenames,dvipsnames]{color}
\usepackage{booktabs}
\usepackage{mathtools}
\usepackage{enumerate}
\usepackage{manfnt}%

\newif\iftodo   % L"a"st \todo-Eintr"age zu ('Baustellen/Unfertiges')
\todotrue
\newif\iftodoshort  % true: Druckt nur den \todo-Marker ohne Kasten aus
\todoshortfalse

\renewcommand{\vec}[1]{\boldsymbol{#1}}
\newcommand{\Reals}{\mathbb{R}}
\newcommand{\Complex}{\mathbb{C}}
\newcommand{\set}[1]{\mathcal{#1}}
\newcommand{\Ex}[1]{\mathbb{E}\left[ {#1} \right]}

\newcommand{\multi}[2]{\genfrac{}{}{0pt}{}{#1}{#2}}

 \newcommand{\diag}[1]{\text{diag}\left( #1 \right)}
\newcommand{\Sn}{\mathbb{S}^{n_t-1}}
\newcommand{\rank}[1]{\text{rank}\left( #1 \right)}

\theoremstyle{definition}
\newtheorem{mydef}{Definition}
\theoremstyle{plain}
\newtheorem{theorem}{Theorem}
\newtheorem{corollary}{Corollary}
\newtheorem{lemma}{Lemma}
\theoremstyle{remark}
\newtheorem{remark}{Remark}

\newlength{\drop}

\newcommand*{\titleUL}{\begingroup
\drop=0.1\textheight
\vspace*{0.5\drop}
\begin{center}
{\LARGE\textsc{$\quad$}}\\[0.5\drop]
{{\LARGE IEEE Transactions on Wireless Communications} \\\vspace*{0.4cm} -- submitted for publication --  }\\[2em]
\rule{\textwidth}{1pt}\par
\vspace{0.5\baselineskip}
{\huge\bfseries Robust Iterative Interference Alignment for \\[.5em] Cellular Networks with Limited Feedback
}\\[0.5\baselineskip]
\rule{\textwidth}{1pt}\par
\vfill
{\Large\textsc{ Jan Schreck$^2$, Gerhard Wunder$^1$, and Peter Jung$^2$}}
\vfill
$^1$Fraunhofer Heinrich Hertz Institute, Einsteinufer 37,
  D-10587 Berlin, Germany\\
$^2$ Technische Universit\"at Berlin
Lehrstuhl f\"ur Informationstheorie und Theoretische Informationstechnik,
Einsteinufer 25, D-10587 Berlin, Germany  
\vfill
\today
\vfill

{\itshape This work was presented in parts at
Globecom Workshop on Emerging Technologies for LTE-Advanced and Beyond-4G.}
\\[1em]
{\itshape This work was partly supported by the German Federal
Ministry of Education and Research (BMBF) under grant 01BU920, the European commission under grant
  FP7-ICT-2011-8, 318555 5GNOW and the Deutsche Forschungsgemeinschaft (DFG)
  under grant JU-2795/2-1.}
\\[1em]
{\itshape \copyright 2014 IEEE. Personal use of this material is
  permitted. Permission from IEEE must be obtained for all other uses,
  in any current or future media, including reprinting/republishing
  this material for advertising or promotional purposes, creating new
  collective works, for resale or redistribution to servers or lists,
  or reuse of any copyrighted component of this work in other works.}
\end{center}
\endgroup}

\begin{document}
\titleUL
\thispagestyle{empty} 
\newpage
 % \title{Robust Iterative Interference Alignment for Cellular Networks with Limited Feedback}
 % \author{Jan Schreck \IEEEmembership{IEEE Member}, Gerhard Wunder,
 % \IEEEmembership{IEEE Member} and Peter Jung \IEEEmembership{IEEE Member} 

 % \thanks{Jan Schreck and Peter Jung are with the Technische Universit\"at Berlin,
 % Lehrstuhl f\"ur Informationstheorie und Theoretische Informationstechnik,
 % Einsteinufer 25, D-10587 Berlin, Germany. Gerhad Wunder is with the Heinrich
 % Hertz Institute, Wireless Communication and Networks Department,  Einsteinufer 37,
 %  D-10587 Berlin, Germany}
% \thanks{This work was partly supported by the German Federal
% Ministry of Education and Research (BMBF) under grant 01BU920, the European commission under grant
%   FP7-ICT-2011-8, 318555 5GNOW and the Deutsche Forschungsgemeinschaft (DFG)
%   under grant JU-2795/2-1.}}

%\maketitle

\begin{abstract}
  In theory coordinated multi-point transmission (CoMP) promises vast
  gains in spectral efficiency. But industrial field trials show
  rather disappointing throughput gains, whereby the major limiting
  factor is proper sharing of channel state information. Many recent
  papers consider this so-called limited feedback problem in the
  context of CoMP.  Usually taking the assumptions: 1) infinite SNR
  regime, 2) no user selection and 3) ideal link adaptation; rendering
  the analysis too optimistic. In this paper we make a step forward
  towards a more realistic assessment of the limited feedback problem
  by introducing an improved metric for the performance evaluation
  which better captures the throughput degradation. We find the
  relevant scaling laws (lower and upper bounds) and show that they
  are different from existing ones. Moreover, we provide a robust
  iterative interference alignment algorithm and corresponding
  feedback strategies achieving the obtained scaling laws. The main
  idea is that instead of sending the complete channel matrix each
  user fixes a receive filter and feeds back a quantized version of
  the effective channel. Finally we underline our findings with
  simulations for the proposed system.
\end{abstract}
\section{Introduction}

Coordinated processing (or so-called coordinated multi-point transmission (CoMP))
of signals by multiple network nodes is a key design element in LTE-A (and
beyond 4G) cellular networks: CoMP algorithms can range from: 1) joint
transmission (fully coherent with message sharing), 2) coordinated beamforming
(without message sharing), to 3) interference coordination (by exchanging e.g.
simple interference indicators). A classical summary of coordination
techniques in multi-cell MIMO cooperative networks can be found in
\cite{Gesbert2010,Marsch2011}. A prominent coordinated beamforming technique
is interference alignment (IA) \cite{Jafar2011} which essentially aligns the
signal space so that multiple interferer appear as a single one.

In theory coherent transmission from multiple base stations to
multiple users promises vast gains in spectral efficiency
\cite{Gesbert2010,Marsch2011}. But, industrial field trials show
rather disappointing throughput gains, whereby the major limiting
factor is proper sharing of channel state information (CSI) and other
overhead among cells \cite{Irmer2011}. Many papers consider the
so-called limited feedback problem. For example, \cite{Caire2010} and
\cite{Kerret2011} considered multiuser MIMO systems and network MIMO
systems, respectively. Reference \cite{Bolcskei2009,Ayach2012}
considered IA for the interference channel. Recently, \cite{Lee2012}
considered IA for the the interfering MAC. All with a focus on the
infinite SNR regime carrying out a system degrees of freedom (DoF)
analysis.

However, even though analytic treatment of the limited feedback
problem has made significant progress in the past, the DoF approach
cannot really account for the throughput degradation experienced in
practice. The main reasons are: 1) The infinite SNR regime where
achieving DoF is optimal is considered. In this regime interference
mitigation instead of signal enhancement is the primary goal. 2) No
user selection is considered, i.e., it is assumed that the optimal
scheduling decision is known. 3) Ideal link adaptation is assumed.
Altogether, this renders the performance analysis too optimistic and
motivates extended analysis of the limited feedback problem.

Now, the question is: How can we get reliable estimates of the performance degradation due to limited feedback. In this paper we take a step forward towards a more realistic answer to this
question. Our approach is universal in the sense that we do not
consider a specific transmit strategy. By considering the
interfering broadcast channel \cite{Suh2008} our results also hold
for the interference channel and the broadcast channel, which are
special cases of the interfering broadcast channel.
In particular we: 
\begin{itemize}
\item introduce an improved metric for the performance evaluation which better
captures the throughput degradation due to limited feedback in practice. The
metric is defined per user instead of sum rate.

\item calculate the rate degradation for any scheduling decision,
any beamforming strategy, and any SNR regime which is a useful
performance benchmark for the design of systems.

\item derive a lower bound on the throughput degradation for IA; 
replacing the too optimistic scaling laws for the number of feedback bits in
the conventional analysis. We prove that the feedback scaling is
$2^{-\frac{B}{2\left(  n_{t}-1\right)  }}$ instead of $2^{-\frac{B}{n_{t}-1}%
}$ in most of the previous work. 

\item introduce a robust iterative IA algorithm with user selection which
achieves the optimal scaling under any SNR regime. The main idea is that
instead of sending the complete channel matrix each user
fixes a receive filter and feeds back a quantized version of
the effective channel. % For this scheme we derive the optimal robust feedback
% decision metric for the terminals.
% Due to the alternating optimization the
% beamformers and receive filters can be locally computed.

\item show that the proposed distributed approach is favorable over 
  centralized approaches in terms of performance, convergence
  speed and computational complexity.

\end{itemize}
We like to disclose that a summary of the results was presented in the
workshop paper \cite{Schreck2013}. In contrast to \cite{Schreck2013}
the paper at hand includes all proofs in detail. Moreover, we develop new
approaches like a partial reverse of Jensen's inequality (Lemma
\ref{lem:lowerBound}).

\textit{Notation:} The inner product of $\vec x\in\mathbb{C}^{N}$ and $\vec y
\in\mathbb{C}^{N}$ is $\langle\vec x , \vec y \rangle= \vec x^{H}\vec y$,
where $\vec x^{H}$ is the conjugate transpose of vector $\vec x$. The vector
$p $-norm is defined as $\| x \|_{p} = (\sum_{i} x_{i}^{p})^{1/p}$. The unit
sphere in $\mathbb{C}^{N}$ is defined as $\mathbb{S}^{N-1}$. The expected
value of a random variable $X$ is $\mathbb{E}\left[  {X} \right]  $.

\section{System Setup}

\label{sec:sysmod}

\subsection{System Model}

Consider the downlink of a cellular network with $K$ base stations,
each equipped with $n_{t}$ transmit antennas, and $U$ user equipments,
each equipped with $n_{r}$ receive antennas. Throughout the paper we
consider an arbitrary but fixed spectral resource element. On this
resource element the channel between base station $b$ and user $m$ is
modeled by the matrix $\vec{H}_{m,b}\in\mathbb{C}^{n_{r}\times
  n_{t}}$ which is constant over one transmission frame and
distributed complex Gaussian with zero mean and unit variance.
%Hence, the coherence time of the
%channel is  assumed to be larger than one transmission interval.
In each transmission frame (time index omitted) all base stations
$b=1,\ldots,K$ select disjoint subsets of users
$\mathcal{S}_{b}\subseteq\mathcal{U}=\{1,\ldots,U\}$ and transmit the signal
$\vec{x}_{b}\in\mathbb{C}^{n_{t}}$. The signal received by user $m\in
\mathcal{\mathcal{S}}_{b}$ is given by
\begin{equation}
y_{m}= \sum_{l=1}^{K}\langle\vec{u}_{m},\vec{H}_{m,l}\vec{x}_{l}\rangle+\langle\vec{u}_{m},\vec{n}_{m}\rangle,
\end{equation}
where $\vec{u}_{m}\in\mathbb{S}^{n_{r}-1}$ is the receive filter and
$\vec{n}_{m} \sim\mathcal{C}\mathcal{N}(0,\vec{I})$
is additive white Gaussian noise. The set of all scheduled users is defined as
the set $\mathcal{S}:= \mathcal{S}_{1} \cup \mathcal{S}_{2}\cup \ldots \cup \mathcal{S}_{K}$ and
the beamforming vectors are given by the function
\begin{equation}
\vec{\pi}:\mathcal{S}\rightarrow\mathbb{S}^{n_{t}-1}.
\end{equation}
Assume that the complex information symbols can be modeled as complex
Gaussian with zero mean and unit variance, $d_{m}\sim \set C \set
N(0,1)$, then the signal transmitted by base station $b$
\begin{equation}
\vec{x}_{b}=\sqrt{\frac{P}{|\mathcal{S}_{b}|}}\sum_{m\in\mathcal{S}_{b}%
}\vec{\pi}(m)d_{m},
\end{equation}
fulfills the average power constrained $\mathbb{E}\left[
  {\Vert\vec{x}_{b}\Vert_{2}^{2}}\right]  =P$, for all $b$. We assume
that each base station distributes its available power
$P_{b}=P$ equally among all users $m\in\mathcal{S}_{b}$. 

Throughout the paper we make the assumption  that
all users $m\in\mathcal{U}$ have perfect knowledge of their own channels $\vec
{H}_{m,l}$, for $l=1,2,\ldots,K$, and we assume no delay in reporting the
CSI, the process of scheduling and the transmission.   

\subsection{Scheduling and Feedback Model}
\label{sec:Sched&FB}
A \textit{scheduling decision} consists of two steps: i) selection of users $\mathcal{S}_b\subseteq\mathcal{U}$ and
ii) computation beamforming vectors 
$\vec{\pi}:\mathcal{S}\rightarrow\mathbb{S}^{n_{t}-1}$. For a given scheduling
decision $(\vec{\pi},\mathcal{S})$ the achievable system sum rate is given by
\begin{equation}
R(\vec{\pi},\mathcal{S};H)=\sum_{l=1}^{K}\sum_{m\in\mathcal{S}_{l}}r_{m}%
(\vec{\pi},\mathcal{S};H),
\end{equation}
where $H=\{\vec{H}_{m,l}:l=1,2,\ldots,K;\,m=1,2,\dots,U\}$ is the list of all
channels. The achievable rate of user $m\in\mathcal{S}_{b}$ is given by the
Shannon rate \footnote{For simplicity of
notation, we assume that with $\mathcal{S}$ also the information about
the cardinality of the partial sets $\mathcal{S}_{1},...,\mathcal{S}_{K}$ is
delivered.}
\begin{equation}
r_{m}(\vec{\pi},\mathcal{S};H)=\max_{\vec{u}\in\mathbb{S}^{n_{t}-1}}
\log\left(  1+\frac{\frac{P}{|\mathcal{S}_{b}|}|\langle\vec{u},\vec{H}
_{m,b}\vec{\pi}(m)\rangle|^{2}}{1+\sum_{l=1}^{K}\sum_{\multi{k\in\mathcal{S}_{l}}{k\neq m}
}\frac{P}{|\mathcal{S}_{l}|}|\langle\vec{u},\vec{H}_{m,l},\vec{\pi}
(k)\rangle|^{2}}\right),
\end{equation}
where the receive filters can be optimized independently by each user;
the receive filter of user $m$ will be denoted by $\vec u_m$. 
In the sequel, we assume that the base stations aim at
maximizing the system sum-rate. Thus, if all base stations have
knowledge of all channels $H$, the optimal scheduling decision $\left(
  \vec{\pi}_{H},\mathcal{S}_{H}\right) $ is the solution to the
optimization problem
\begin{equation}
\max_{\mathcal{S}\subseteq\mathcal{U}}\max_{\vec{\pi}:\mathcal{S}%
\rightarrow\mathbb{S}^{n_{t}-1}}R(\vec{\pi},\mathcal{S};H).
\label{eq:sched_perfectCSIT}%
\end{equation}
Therefore, the optimal system sum rate is $R(\vec{\pi}_H,\mathcal{S}_H;H)$. 

In the following we assume that the base stations collect quantized
CSI through a rate-constrained feedback channel.  For fixed receive
filters $\vec{u}_{k}$ each user $k\in\mathcal{U}$ quantizes and feeds
back the effective channels
$\vec{\hat{h}}_{k,l}:=(\vec{H}_{k,l})^{H}\vec{u}_{k}$ to all base
stations $l=1,\ldots,K$. In particular user $k\in\mathcal{U}$ uses
random vector quantization (RVQ) on the normalized
effective channels
\begin{equation} 
\vec{h}_{k,l}:=\frac{\vec{\hat{h}}_{k,l}}{\Vert\vec{\hat{h}}_{k,l}\Vert_{2}%
},\quad\forall\,l\in [1,K].
\end{equation}
The normalized effective
channels are quantized using a random codebook $\mathcal{V}_{k}\subset
\mathbb{S}^{n_{t}-1}$, with $2^{B}$ isotropically distributed
elements. Each user uses an independent copy of the random
codebook which ensures that the feedback messages from
different users are linearly independent, almost surely.  Each user
$k$ feeds back the indices of the elements
\begin{equation}
  \label{eq:robFB}
  \vec{v}_{k,l}:=\underset{\vec{v}\in\mathcal{V}_{k}%
  }{\arg\min}\,\left(
    1-|\langle\vec{h}_{k,l},\vec{v}\rangle|^{2}\right), \quad
  \forall\, l\in [1,K], 
\end{equation}
to all base stations. Here,
$1-|\langle\vec{h}_{k,l},\vec{v}\rangle|^{2}$ is the squared chordal
distance and
$\min_{\vec{v}\in\mathcal{V}_k}(1-|\langle\vec{h}_{k,l},\vec{v}\rangle|^{2})$
is the quantization error.  Later on in Section
\ref{sec:IAana} we will show that the chordal distance is a reasonable
and robust quantization metric for the considered
systems. Equivalently, the quantization problem can be formulated on
the complex Grassmann manifold $\set G(n_t,1)$ which is the set of
all one dimensional subspaces of $\Complex^{n_t}$ (see
e.g. \cite{Love2003} for further details).  To simplify our analysis
we assume that the channel norm
$\mu_{k,l}:=\Vert\vec{\hat{h}}_{k,l}\Vert_{2}$ is perfectly known to
all base stations.

After receiving the feedback messages from all users
$k\in \set U$, each base
station $l=1,...,K$, has knowledge of the quantized effective channels
\begin{equation}
V:= \{ \vec{\hat{v}}_{k,l} =\mu_{k,l}\vec{v}_{k,l} :
k\in \set U, l\in [1,K] \}.
\label{eq:partCSI}%
\end{equation}
Based on quantized CSI $V$ the scheduling decision $\left(
  \vec{\pi}_{V},\mathcal{S}_{V}\right) $ is found by solving the
problem
\begin{equation}
\max_{\mathcal{S}\subseteq\mathcal{U}}\max_{\vec{\pi}:\mathcal{S}%
\rightarrow\mathbb{S}^{n_{t}-1}}R\left(  \vec{\pi},\mathcal{S};V\right),
\label{eq:partCSI_sched}%
\end{equation}
instead of problem \eqref{eq:sched_perfectCSIT}. 
\begin{remark}
  In general the scheduling decisions with quantized and perfect CSI
  are not equal, $\left(\vec{\pi}_{V},\mathcal{S}_{V}\right)\neq
  \left( \vec{\pi}_{H},\mathcal{S}_{H}\right)$. Therefore, the
  achievable sum rate with quantized CSI is smaller or equal the
  achievable sum rate with perfect CSI,
  $R(\vec{\pi}_V,\mathcal{S}_V;H) \leq
  R(\vec{\pi}_H,\mathcal{S}_H;H)$.
\end{remark}

The core of this paper is to explore the performance degradation under
the scheduling decisions based on quantized CSI $V$ (Section
\ref{sec:Ana}) under suitable algorithms for the optimization problems
\eqref{eq:sched_perfectCSIT} and \eqref{eq:partCSI_sched} which we
discuss in Section \ref{sec:IA}.

\section{Rate Loss Gap Analysis}
\label{sec:Ana} 

\subsection{Known Results}
\label{sec:known}
In the literature usually the rate gap $r_{m}(\vec{\pi
}_{H},\mathcal{U},H)-r_{m}(\vec{\pi}_{V},\mathcal{U},H)$ is
analyzed. That is, the set of active users is fixed and perfect link
adaptation is assumed. Moreover, most papers consider a specific
system setup (e.g. the broadcast channel or the $K$-user interference
channel) and a specific beamforming strategy (e.g. zero forcing
beamforming or IA). Based on these assumptions, the influence of
quantized CSI on the sum rates or user rates is analyzed. Let us
shortly summarize some of the results.

In \cite{Caire2010,Jindal2006} a broadcast channel with
$|\mathcal{U}|=n_{t}$ single antenna users and zero forcing
beamforming is assumed. Let us denote $\vec{\pi}_{\text{ZF},H}$ and
$\vec{\pi}_{\text{ZF},V}$ as the zero forcing beamforming solutions
with perfect and quantized CSI, respectively.  According to
\cite{Jindal2006}, limited feedback with $B$ feedback bits per user
incurs a throughput loss relative to zero forcing with perfect CSI
bounded by
\begin{equation}
\mathbb{E}\left[  {r_{m}(\vec{\pi}_{\text{ZF},H},\mathcal{U},H)-r_{m}(\vec
{\pi}_{\text{ZF},V},\mathcal{U},H)}\right]  <\log(1+P2^{-\frac{B}{n_{t}-1}}).
\label{eq:ZFscal}%
\end{equation}
This result has been recently generalized in \cite{Kerret2011} for
network MIMO.

For IA, a limited feedback scheme for the $K$-user
interference channel is proposed in \cite{Rezaee2012}. Quantization is based
on Grassmannian representation of the channel matrices. The throughput
loss due to the channel quantization scales like
\begin{equation}
  \label{eq:scaleIA}
  \mathbb{E}\left[  {r_{m}(\vec{\pi}_{\text{IA},H},\mathcal{U},H)-r_{m}(\vec
{\pi}_{\text{IA},V},\mathcal{U},H)}\right] < \log\left(1+P2^{-\frac{B}{N_{g}}+1}\right),
\end{equation}
where $N_{g}=2n_{r}((K-1)n_{t}-n_{r})$ is the real dimension of the
Grassmannian manifold.  In \cite{Krishnamachari2010} similar results
have been obtained for a system using OFDM. An in depth treatment of
the scaling law analysis in Grassmannian manifolds can be found in
\cite{Krishnamachari2011}.

In \cite{Lee2012} a cellular system with two base stations and four
users was considered and it was shown that the rate gap scales exactly
like \eqref{eq:ZFscal}. As we will see this result can not be
generalized to systems with more than two cells. 

In the following  we show that the scaling law \eqref{eq:ZFscal} 
is to optimistic, if we consider more general systems and a slightly
different but more realistic metric. Moreover we will see that the
scaling laws \eqref{eq:scaleIA} can be significantly improved if we use a
different feedback and IA strategy.  

\subsection{An Improved Metric}
In this paper we assume that the CSI is used by the base stations to
perform (i) beamforming, (ii) scheduling, and (iii) link adaptation.  If
the base stations have only quantized CSI, each of these tasks causes a
rate loss compared to the performance with perfect CSI.
Therefore, we define the following per
user performance metric
\begin{align}
\Delta r_{m}(\vec{\pi}_{H},\vec{\pi}_{V})  & =\max\{r_{m}(\vec{\pi}%
_{H},\mathcal{S}_{H};H)-r_{m}(\vec{\pi}_{V},\mathcal{S}_{V};H),r_{m}(\vec{\pi
}_{H},\mathcal{S}_{H};H)-r_{m}(\vec{\pi}_{V},\mathcal{S}_{V}%
;V)\} \nonumber\\
&  =r_{m}(\vec{\pi}_{H},\mathcal{S}_{H};H)-\min\{r_{m}(\vec{\pi}%
_{V},\mathcal{S}_{V};H),r_{m}(\vec{\pi}_{V},\mathcal{S}_{V};V)\}\label{eq:improved_metric} .
\end{align}
Because of the per user formulation $\Delta
r_{m}(\vec{\pi}_{H},\vec{\pi}_{V})$ is not necessarily positive for
all $H$. The rate loss gap $\Delta r_{m}(\vec{\pi}_{H},\vec{\pi}_{V})$ captures
the following effects:
\begin{enumerate}
\item If $r_{m}(\vec{\pi}_{V},\mathcal{S}_{V};H) >
  r_{m}(\vec{\pi}_{V},\mathcal{S}_{V};V)$ the rate gap is 
  $r_{m}(\vec{\pi}_{H},\mathcal{S}_{H};H)-r_{m}(\vec{\pi}_{V},\mathcal{S}_{V};V)$. 
  Thus, the rate gap
  captures the rate loss due to beamforming, scheduling and link
  adaptation based on quantized CSI, since it is assumed that the base
  station transmits with a rate $r_{m}(\vec{\pi}_{V},\mathcal{S}_{V};V)$. 
\item If $r_{m}(\vec{\pi}_{V},\mathcal{S}_{V};H) <
  r_{m}(\vec{\pi}_{V},\mathcal{S}_{V};V)$, the rate gap is
  $r_{m}(\vec{\pi}_{H},\mathcal{S}_{H};H)-r_{m}(\vec{\pi}_{V},\mathcal{S}
  _{V};H)$ and describes the rate loss due to beamforming and
  scheduling based on quantized CSI. We do not consider link
  adaptation, because even if allocation of a rate
  $r_{m}(\vec{\pi}_{V},\mathcal{S}_{V};V)>r_{m}(\vec{\pi}_{V},\mathcal{S}_{V};H)$
  causes an outage event with high probability in practice mechanisms
  like automatic repeat requests are used to handle such events.
\end{enumerate}
As we will see, $\Delta r_{m}(\vec{\pi}_{H},\vec{\pi}_{V})$ is
strong enough to address some of the drawbacks of the conventional
analysis (summarized in Subsection \ref{sec:known}) and leads to
indeed different results.

In the remainder, we will derive lower and upper bounds on
$\Delta r_{m}(\vec{\pi}_{H},\vec{\pi}_{V})$ for symmetric systems which are
defined as follows. 
\begin{mydef} 
  In a {\it symmetric system} the random channels $\vec H_{m,l}$ are
  independent and identically distributed for all $m\in \set U$ and
  all $l=1,\ldots,K$.  Further, the distribution of the effective
  channels $(\vec H_{m,l})^H\vec u_m$, for all $l=1,\ldots,K$ and each
  user $m\in \set U$ given some fixed arbitrary receive filter $\vec
  u_m$, is the same and isotropic.
\end{mydef}
The following
lemma sets the basis for our analysis; it allows us to bound the rate gap $\Delta
r_{m}(\vec{\pi}_{H},\vec{\pi}_{V})$ in terms of the \emph{same}
scheduling decisions.% $\left( \pi,\mathcal{S}\right)$. 
\begin{lemma}
  \label{lem:motivation} Let $\mathcal{S}_{H}$ and
  $\vec{\pi}_{H}:\mathcal{S}%
  _{H}\rightarrow\mathbb{S}^{n_{t}-1}$ denote the optimal user
  selection and the optimal beamforming vectors under perfect CSI $H$
  according to \eqref{eq:sched_perfectCSIT}. Similarly, let
  $\mathcal{S}_{V}$ and $\vec{\pi
  }_{V}:\mathcal{S}_{V}\rightarrow\mathbb{S}^{n_{t}-1}$ be the optimal
  user selection and the optimal beamforming vectors under quantized
  CSI $V$ according to \eqref{eq:partCSI_sched}. Assume a symmetric
  system and fix some arbitrary user $m\in\mathcal{U}$, then the
  expected rate gap $\mathbb{E}\left[ {\Delta
      r_{m}(\vec{\pi}_{H},\vec{\pi}_{V})}\right] $ is bounded by
\begin{multline}
\mathbb{E}\left[  {\max\left\{  r_{m}(\vec{\pi}_{V},\mathcal{S}_{V}%
;H)-r_{m}(\vec{\pi}_{V},\mathcal{S}_{V};V),0\right\}  }\right] \\
\leq\mathbb{E}\left[  {\Delta r_{m}(\vec{\pi}_{H},\vec{\pi}_{V})}\right] \\
\leq3\,\mathbb{E}\left[ \max_{\mathcal{S}\subseteq U}\max_{\vec{\pi}:\set S
  \rightarrow \mathbb S^{n_t-1}} {\left\vert
r_{m}(\vec{\pi},\mathcal{S};H)-r_{m}(\vec{\pi},\mathcal{S};V)\right\vert
}\right].
\end{multline}

\end{lemma}

\begin{IEEEproof}
  First we need to show that $\mathbb{E}\left[ {r_{m}(\vec{\pi}_{H}%
      ,\mathcal{S}_{H};V)}\right] \leq\mathbb{E}\left[
    {r_{m}(\vec{\pi}%
      _{V},\mathcal{S}_{V};V)}\right] $ which does not trivially
  follow from the sum rate maximization \eqref{eq:partCSI_sched}. To
  see this assume that $\mathbb{E}\left[
    {r_{m}(\vec{\pi}_{H},\mathcal{S}_{H};V)}\right] >\mathbb{E}\left[
    {r_{m}(\vec{\pi}_{V},\mathcal{S}_{V};V)}\right] $ for some user
  $m$. Since $\mathbb{E}\left[
    {r_{m}(\vec{\pi}_{H},\mathcal{S}_{H};V)}\right] =\mathbb{E}\left[
    {r_{l}(\vec{\pi}_{H},\mathcal{S}_{H};V)}\right] $ and
  $\mathbb{E}\left[ {r_{m}(\vec{\pi}_{V},\mathcal{S}_{V};V)}\right]
  =\mathbb{E}\left[ {r_{l}(\vec{\pi}_{V},\mathcal{S}_{V};V)}\right]
  $ when $m\neq l$, it follows from the symmetry of the system that
  $\mathbb{E}\left[ {r_{m}%
      (\vec{\pi}_{H},\mathcal{S}_{H};V)}\right] >\mathbb{E}\left[
    {r_{m}(\vec{\pi }_{V},\mathcal{S}_{V};V)}\right] $ for all
  $m\in\mathcal{U}$. Hence, we have
\begin{equation}
\mathbb{E}\left[  \sum_{m\in\mathcal{U}}r_{m}(\vec{\pi}_{V},\mathcal{S}%
_{V};V)\right]  <\mathbb{E}\left[  \sum_{m\in\mathcal{U}}r_{m}(\vec{\pi}%
_{H},\mathcal{S}_{H};V)\right]  ,
\end{equation}
which contradicts with the definition of $\mathcal{S}_{V}$ and $\vec{\pi
}_{V}$ given in \eqref{eq:partCSI_sched}. Therefore, 
\begin{equation}
  \label{eq:rmV}
\mathbb{E}\left[  {r_{m}(\vec{\pi}_{H},\mathcal{S}_{H};V)}\right]
\leq\mathbb{E}\left[  {r_{m}(\vec{\pi}_{V},\mathcal{S}_{V};V)}\right]  ,  
\end{equation}
must hold for all $m\in \set U$. In a similar manner we can show that
\begin{equation}
  \label{eq:rmH}
\mathbb{E}\left[  {r_{m}(\vec{\pi}_{V},\mathcal{S}_{V};H)}\right]
\leq\mathbb{E}\left[  {r_{m}(\vec{\pi}_{H},\mathcal{S}_{H};H)}\right]
\end{equation}
holds for all $m\in\mathcal{U}$.  Inequalities \eqref{eq:rmV} and
\eqref{eq:rmH} state that in expectation the sum-rate optimal
scheduling decision maximizes also the individual per user rates.
To prove the upper bound we write $\mathbb{E}\left[  {\Delta r_{m}(\vec{\pi}_{H},\vec{\pi}_{V})}\right]$ as 
\begin{multline}
\mathbb{E}\left[  {\Delta r_{m}(\vec{\pi}_{H},\vec{\pi}_{V})}\right]
=\\ \mathbb{E}[r_{m}(\vec{\pi}_{H},\mathcal{S}_{H};H)-r_{m}(\vec{\pi}%
_{V},\mathcal{S}_{V};H)+\max\{r_{m}(\vec{\pi}_{V},\mathcal{S}_{V}%
;H)-r_{m}(\vec{\pi}_{V},\mathcal{S}_{V};V),0\}].
\end{multline}
The first term can be bounded by
\begin{align}
&  \mathbb{E}[r_{m}(\vec{\pi}_{H},\mathcal{S}_{H};H)-r_{m}(\vec{\pi}%
_{V},\mathcal{S}_{V};H)]\nonumber\\
&  =\mathbb{E}[r_{m}(\vec{\pi}_{H},\mathcal{S}_{H};H)-r_{m}(\vec{\pi}%
_{V},\mathcal{S}_{V};V)+r_{m}(\vec{\pi}_{V},\mathcal{S}_{V};V)-r_{m}(\vec{\pi
}_{V},\mathcal{S}_{V};H)]\label{eq:firsteq}\\
&  \leq\mathbb{E}[r_{m}(\vec{\pi}_{H},\mathcal{S}_{H};H)-r_{m}(\vec{\pi}%
_{H},\mathcal{S}_{H};V)+r_{m}(\vec{\pi}_{V},\mathcal{S}_{V};V)-r_{m}(\vec{\pi
}_{V},\mathcal{S}_{V};H)]\label{eq:firstineq}\\
&  \leq2\mathbb{E}[\max_{\mathcal{S}\subseteq U}\max_{\vec{\pi}:\set S
  \rightarrow \mathbb S^{n_t-1}}\left\vert
r_{m}(\vec{\pi},\mathcal{S};H)-r_{m}(\vec{\pi},\mathcal{S};V)\right\vert ]\nonumber. 
\end{align}
Equation \eqref{eq:firsteq} holds since we simply added a $0 = -
r_{m}(\vec{\pi}%
_{V},\mathcal{S}_{V};V)+r_{m}(\vec{\pi}_{V},\mathcal{S}_{V};V)$. The
first inequality \eqref{eq:firstineq} holds according to
\eqref{eq:rmV}. Further by \eqref{eq:rmH} we have $\mathbb{E}[r_{m}(\vec{\pi}_{H},\mathcal{S}_{H};H)-r_{m}(\vec{\pi}%
_{V},\mathcal{S}_{V};H)]\geq 0$. Since
\begin{equation}
\mathbb{E}[\max\{r_{m}(\vec{\pi}_{H},\mathcal{S}_{H};H)-r_{m}(\vec{\pi}%
_{V},\mathcal{S}_{V};V),0\}]\leq\mathbb{E}[\max_{\mathcal{S}\subseteq U}\max_{\vec{\pi}:\set S
  \rightarrow \mathbb S^{n_t-1}}\left\vert r_{m}(\vec{\pi},\mathcal{S};H)-r_{m}(\vec{\pi
},\mathcal{S};V)\right\vert ]. 
\end{equation}
is also true the upper bounds follows. Since according to \eqref{eq:rmH}
\begin{equation}
\mathbb{E}[r_{m}(\vec{\pi}_{H},\mathcal{S}_{H};H)-r_{m}(\vec{\pi}%
_{V},\mathcal{S}_{V};H)]\geq0, 
\end{equation}
the lower bound follows by setting the first term in \eqref{eq:improved_metric} equal to $0$. 

\end{IEEEproof}
We offer some brief remarks. 
\begin{remark}
  Lemma \ref{lem:motivation} is tight if $H=V$. On the other hand, if
  $H$ and $V$ are not related the bound can be arbitrary loose.
  However, since we assume that $V$ is a good approximation of $H$ the
  bounds in Lemma \ref{lem:motivation} can be assumed to be reasonably
  tight.
\end{remark}

\begin{remark}
  Even though we assumed achievable rates $\log(1+x)$, it is possible
  to consider extensions of this lemma which incorporate more general
  utility functions. In addition, the assumptions on the channel
  distribution may be relaxed here but they are required in the
  subsequent theorems.
\end{remark}

\subsection{Main Result}

\label{sec:arbitBeam} 

The main result in this subsection is an upper bound on the expected
rate gap $\Ex{\Delta r(\vec \pi_H , \vec \pi_V)}$ defined in
(\ref{eq:improved_metric}). In contrast to previous results
(summarized in Subsection \ref{sec:known}), the following theorem
holds for any receive and transmit strategy, incorporates user
selection and is valid for any SNR regime.
\begin{theorem}
\label{theo:scalingANY} 
Assume a symmetric system with limited feedback. Let the transmit
beamformer $\vec \pi$ and the user selection $\set S$ be arbitrary but
fixed. If each
user $m\in\mathcal{U}$ uses RVQ \eqref{eq:robFB} with $B$ bits per base station feedback link,
\begin{multline}
  \mathbb{E}\left[  |r_{m}(\vec{\pi},\mathcal{S};H)-r_{m}(\vec{\pi
},\mathcal{S};V)|\right] \\
  \leq2\log\left(  1+\frac{P}{2}\left[  K\left(  K-1\right)  \mathbb{E}%
\left[  \mu_{m,1}^{4}\right]  2^{-\frac{B}{n_{t}-1}}+2^{-\frac{B}{2(n_{t}-1)}%
}\left(  \mathbb{E}^{\frac{1}{4}}\left[  \mu_{m,1}^{8}\right]  \mathbb{+}%
K\mathbb{E}^{\frac{1}{2}}\left[  \mu_{m,1}^{2}\right]  \right)  \right]
\right)  \label{eq:theo1},
\end{multline}
holds for any SNR $P$. 
\end{theorem}
The proof is presented in Appendix \ref{proof:scalingANY}. Theorem
\ref{theo:scalingANY} holds regardless of the transmit strategy, i.e.,
for any beamforming and user selection strategy. None of the known
results presented in Section \ref{sec:known} hold with such
generality.

\begin{remark}
  In contrast to the known results \eqref{eq:ZFscal} and
  \eqref{eq:scaleIA} we obtain a different scaling for the RVQ scheme
  \eqref{eq:robFB}. It is significantly better than the result
  \eqref{eq:scaleIA} for IA with Grassmannian feedback, but
  requires two times more feedback bits than the result
  \eqref{eq:ZFscal} for the broadcast channel with zero forcing.
\end{remark}

\begin{remark}
  Assuming the worst-case decision for each transmit beamformer (given
  fixed receive filters), the upper bound \eqref{eq:cor1Tight} in the
  proof of Theorem \ref{theo:scalingANY} is tight. Therefore, also the
  upper bound in Lemma \ref{lem:motivation} is tight. Hence, using
  Lemma \ref{lem:ExChor} it follows that the minimum chordal distance
  is a robust feedback metric for the considered systems.
\end{remark}
In the next
section we present an interference alignment algorithm and derive a
corollary which adapts Theorem \ref{theo:scalingANY} to this algorithm.

\section{Interference Alignment with Quantized CSI and User Selection}

\label{sec:IA}
Without requiring further constraints, scheduling  problem
\eqref{eq:sched_perfectCSIT} and \eqref{eq:partCSI_sched} are NP-hard
\cite{Weeraddana2011}. Therefore, good sub-optimal solutions are
required.  In this section we present an algorithm that efficiently
solves the scheduling problem by alternating the optimization of
receive filters and transmit beamformers. As we will see, the
algorithm is robust to CSI quantization and keeps the
feedback overhead low. 

\subsection{Cellular Interference Alignment}
The IA algorithm presented below uses concepts of spatial IA and user
selection; it aims on finding beamforming vectors $\vec{\pi}_{\text{IA}}$ and user
sets $\mathcal{S}_{1},...,\mathcal{S}_{K}$ such that the following conditions
hold
\begin{align}
|\langle\vec{u}_{m},\vec{H}_{m,b}\vec{\pi}_{\text{IA}}(m)\rangle|^{2}  &  \geq
c_{0},\quad\forall\,b=1,2,\ldots,K\text{ and }m\in\mathcal{S}_{b}
\label{eq:IAgain}\\
|\langle\vec{u}_{m},\vec{H}_{m,l}\vec{\pi}_{\text{IA}}(k)\rangle|^{2}  &
=0,\quad\forall\,b,l=1,2\ldots,K\text{ and }k\in\mathcal{S}_{l},\,m\in
\mathcal{S}_{b}\text{ with }k\neq m \label{eq:IAzero}
\end{align}
where $c_0>0$ is a positive constant.
Condition \eqref{eq:IAgain} states that for each active user the desired
effective channels are non zero and condition \eqref{eq:IAzero} states that
all interfering channels are zero. Typically, IA is analyzed using the concept of
DoF \cite{Cadambe2008} which are defined as
\begin{equation}
  \label{eq:IAS}
  d =\lim_{P\rightarrow\infty}\frac{R\left(  \vec{\pi}%
,\mathcal{S};H\right)  }{\log(P)}+o(\log(P)).
\end{equation}
Intuitively, the DoF can be seen as the number of interference-free parallel
data streams that can be transmitted simultaneously in a network. 
The DoFs for symmetric cellular networks with spatial interference
alignment have been analyzed in \cite{Schreck2012WSA}, where we have
shown that condition \eqref{eq:IAgain} and \eqref{eq:IAzero} can be
fulfilled, almost surely, if
\begin{equation}
n_{t}\geq\frac{SK+1}{2}, \label{eq:feasible}%
\end{equation}
with $n_{r}=n_{t}$, $S=|\mathcal{S}_{b}|=|\mathcal{S}_{l}|$, for all
$l,b=1,2,\ldots,K$, and a single data stream per user. Hence, spatial IA is
feasible, almost surely, if the number of active users per base station is bounded by
\begin{equation}
|\mathcal{S}_{b}|\leq\frac{1}{K}\left(  {2n_{t}}-1\right)  ,\text{ }%
b=1,\ldots,K. \label{eq:maxUE}%
\end{equation}
% Figure \ref{fig:feas} depicts the feasible DoF for $n_{t}=5$ and a different
% number of base stations $T$ and active users $K=|\mathcal{S}_b|$, for all $b=1,\ldots,T$.
% \begin{figure}[htb]
% \centering
% \includegraphics[width =.5\linewidth]{fig/feas.eps}\caption{Feasibility of
% interference alignment for a symmetric cellular network with $B$ base
% stations, $K$ users per base station, $d$ data streams per user and $5$
% antennas per node.}%
% \label{fig:feas}%
% \end{figure}In the following we present two spatial IA algorithms tailored to 
% cellular networks. The algorithms were first proposed by the authors in
% \cite{Schreck2011WSA} and extended with user selection in
% \cite{Schreck2012ICASSP}.

\subsection{Minimum Interference Algorithm}

\label{sec:Alg} The algorithm presented here aims on minimizing
interference and may achieves interference alignment. In the sequel
we will call this algorithm minimum interference algorithm. The
minimum interference algorithm with user selection is summarized in
Algorithm \ref{alg1}. To ensure that all interference can be canceled,
the maximum number of active users is selected according to the
feasibility condition \eqref{eq:maxUE}. At the beginning of each
transmission frame the active users are selected according to some
metric, e.g, maximum fairness, maximum channel gain or other
requirements, possibly defined by higher layers. Having determined the
set of active users the alternating optimization of receive filters
and beamformers is performed. 

Many algorithms that achieve interference alignment or other related
objectives have been proposed in the literature, e.g.,
\cite{Gomadam2008,Suh2008,Schreck2011WSA,Schreck2012WSA}. To our best
knowledge we are the first to propose that the quantized CSI is given
by the quantized effective channel $V$ defined in \eqref{eq:partCSI}.  

Even if we only consider single stream transmissions, extensions to
multi-stream transmissions are straightforward.  Multi-stream
transmission requires that each user feeds back the effective channel
for all streams that it wants to transmit.  Based on this additional
information additional streams can be treated like additional users.

\begin{algorithm}
[htb] \caption{minimum interference algorithm with user selection} \label{alg1}

\begin{algorithmic}
\State\textbf{Begin of transmission frame}:

\State Transmit common pilots to all users and make an estimate $\vec{H}%
_{m,b}$, with $b\in [1,K]$ and $m\in\mathcal{U}$.

\State For $b\in [1,K]$ select $\mathcal{S}_{b}\subseteq\mathcal{U}$
according to \eqref{eq:maxUE} by central control. \State Set $\vec{\pi
}(k)=1/\sqrt{n_{t}}(1,1,\ldots,1)^{T}$ for all $k\in\{\mathcal{S}%
_{1},\mathcal{S}_{2},\ldots,\mathcal{S}_{K}\}$. \Repeat\State Transmit
dedicated pilots. \For{$b=1,2,\ldots,K$} \State Compute receive filter matrix
$\vec{u}_{k}$, for all $k\in\mathcal{S}_{b}$, according to \eqref{eq:rx}.
\State Quantize and feed back the effective channels $\vec{\hat{v}}_{k,l}$,
for all $k\in\mathcal{S}_{b}$, $l=1,\ldots,K$, according to
\eqref{eq:partCSI}. \EndFor\For{$b=1,2,\ldots,K$} \State Compute beamforming
vectors $\vec{\pi}(k)$, for all $k\in\mathcal{S}_{b}$, according to
\eqref{eq:txfinal}. \EndFor

\Until{termination condition is satisfied (e.g. maximum number of iterations,
minimum residual interference, \ldots).}

\State\textbf{End of transmission frame}
\end{algorithmic}
\end{algorithm}

\subsubsection{Receive filter optimization based on perfect CSI}

\label{sec:rxOpt} At the beginning of each transmission frame, orthogonal
common pilots are transmitted, so that, all users $k$ can measure the channel
matrices $\vec{H}_{k,b}$, for all $b$. Common pilots are necessary to compute
the effective channels at the terminals and must be retransmitted in intervals
depending on the
coherence time of the channel. 

During the receive filter
optimization all transmit beamformers $\vec{\pi}$ are fixed. In the first iteration the beamforming vectors are set to $\vec{\pi
}(k)=1/\sqrt{n_{t}}(1,1,\ldots,1)^{T}$.  Based
on dedicated (precoded) pilots each user $k\in\mathcal{S}_{b}$ measures the effective channels $\vec{H}_{k,l}\vec{\pi}(m),l=1,\ldots
,K,l\neq b,m\in\mathcal{S}_{l}$. Based on the measured channels, the receive
filter of user $k\in\mathcal{S}_{b}$ is given by
\begin{equation}
\vec{u}_{k}=\nu_{\min}\bigl(\vec{\Theta}_{b,k}\bigr), \label{eq:rx}%
\end{equation}
where $\nu_{\min}(\vec{X})$ is defined as the eigenvector corresponding to the
smallest eigenvalue of the Hermitian matrix $\vec{X}$. The out of cell
interference covariance matrix $\vec{\Theta}_{b,k}$ is defined as
\begin{equation}
\vec{\Theta}_{b,k}=\sum_{
\multi{l=1}{l\neq b}%
}^{K}\sum_{m\in\mathcal{S}_{l}}\vec{H}_{k,l}\vec{\pi}(m)\bigl(\vec{H}%
_{k,l}\vec{\pi}(m)\bigr)^{H}. \label{eq:Theta}%
\end{equation}
In the receive filter optimization no intra-cell interference is
considered. Intra-cell interference is considered in the beamformer
optimization only. This approach ensures that the intra-cell interference
gets aligned with the out-of-cell interference.

\subsubsection{Transmit beamformer optimization based on quantized CSI}

\label{sec:txOpt} The transmit beamformer optimization is performed at the base
stations and is based on quantized CSI $V$ \eqref{eq:partCSI}. 
 The transmit beamformers are computed in two steps. 
First, the transmit subspace
which causes minimum out-of-cell interference is determined. Second, the
intra-cell interference is canceled by a zero forcing step. Consider the
reciprocal network. The reciprocal precoded channel from user $k$ in cell $b$
to base station $l$ is given by $\overleftarrow{\vec{v}}_{k,l}=\vec{\hat{v}%
}_{k,l}$. For base station $b$ the transmit subspace which causes minimum
out-of-cell interference is given by
\begin{equation}
\vec{\Pi}_{b}=\nu_{\min}^{|\mathcal{S}_{b}|}\bigl(  \overleftarrow{\vec
{\Theta}}_{b}\bigr)  \in\mathbb{C}^{n_{t}\times|\mathcal{S}_{b}|},
\label{eq:tx}
\end{equation}
where $\nu_{\min}^{N}(\vec{X})$ is defined as the eigenvectors corresponding
to the $N$ smallest magnitude eigenvalues of the Hermitian matrix $\vec{X}$. 
The interference covariance matrix $\overleftarrow{\vec{\Theta}}_{b}$ is
defined as
\begin{equation}
\overleftarrow{\vec{\Theta}}_{b}=\sum_{\multi{l=1}{l\neq b}
}^{K}\sum_{m\in\mathcal{S}_{l}}\overleftarrow{\vec{v}}_{m,b}(\overleftarrow
{\vec{v}}_{m,b})^{H}.
\end{equation}
Finally, the intra-cell interference is canceled by an additional zero forcing
step. The zero forcing beamformer $\vec{w}_{m,b}\in\mathbb{C}^{|\mathcal{S}%
_{b}|}$ for user $m$ in cell $b$ is chosen from the null space of the
effective channels $\vec{v}_{k,b}^{H}\vec{\Pi}_{b}$, with $k\in\mathcal{S}%
_{b}$ and $k\neq m$, such that, the transmit beamformer for user $m\in
\mathcal{S}_{b}$ is given by
\begin{equation}
\vec{\pi}(m)=\vec{\Pi}_{b}\vec{w}_{m,b}. \label{eq:txfinal}%
\end{equation}

\subsubsection{Convergence of the residual interference}
The convergence of the residual interference using the minimum interference algorithm can be
proved in a similar manner as the proof of convergence in \cite{Gomadam2008}
for the $K$-user interference channel. Key observations are the
following. First, the out-of-cell interference is monotonically decreased when
computing the receive filter \eqref{eq:rx}. Second, the transmit beamformer
computations \eqref{eq:txfinal} decreases the out-of-cell interference and
nulls all intra-cell interference.

Note that even if the residual interference converges to a local
minimum, it is not guaranteed that the algorithm converges to a unique
solution.

\subsection{Rate Loss Gap Analysis}
\label{sec:IAana}
Together with Lemma
\ref{lem:motivation} we have the following corollary which tailors Theorem
\ref{theo:scalingANY} to the minimum interference algorithm.
\begin{corollary}
\label{cor:scalingALG} Under the assumptions of Theorem \ref{theo:scalingANY},
in any iteration of Algorithm \ref{alg1} the average
rate loss per user is upper bounded by
\begin{multline}
\mathbb{E}\left[  \Delta r_{m}(\vec{\pi}_{H},\vec{\pi}_{V})\right]
\leq 6 \log\left( \frac{P}{2}\left[K^{2}n_{r}^{2}2^{-\frac{B}{n_{t}-1}}+(n_{r}+K)2^{-\frac{B}{2\left(
n_{t}-1\right)  }}\right] \right)   \\ \leq 3P\left(  K^{2}n_{r}^{2}2^{-\frac{B}{n_{t}-1}}+(n_{r}+K)2^{-\frac{B}{2\left(
n_{t}-1\right)  }}\right).  \label{eq:cor2}
\end{multline}
\end{corollary}
\begin{IEEEproof}
Using Lemma \ref{lem:motivation} we have
\begin{equation}
  \label{eq:cor:1}
\mathbb{E}\left[  \Delta r_{m}(\vec{\pi}_{H},\vec{\pi}_{V})\right]
\leq3\,\mathbb{E}\left[ \max_{\mathcal{S}\subseteq U}\max_{\vec{\pi}:\set S
  \rightarrow \mathbb S^{n_t-1}} {\left\vert
r_{m}(\vec{\pi},\mathcal{S};H)-r_{m}(\vec{\pi},\mathcal{S};V)\right\vert
}\right].  
\end{equation}
Now, we can use Theorem \ref{theo:scalingANY} and obtain 
\begin{multline}
\mathbb{E}\left[  \Delta r_{m}(\vec{\pi}_{H},\vec{\pi}_{V})\right]
\\ \leq 6 \log\left(  1+\frac{P}{2}\left[  K\left(  K-1\right)  \mathbb{E}%
\left[  \mu_{m,1}^{4}\right]  2^{-\frac{B}{n_{t}-1}}+2^{-\frac{B}{2(n_{t}-1)}%
}\left(  \mathbb{E}^{\frac{1}{4}}\left[  \mu_{m,1}^{8}\right]  \mathbb{+}%
K\mathbb{E}^{\frac{1}{2}}\left[  \mu_{m,1}^{2}\right]  \right)  \right]
\right)
\end{multline}
When optimizing the beamformers using Algorithm \ref{alg1} all receive
filters are fixed. Thus, we can compute the expected values
\begin{align*}
  \mathbb{E} \left[  \mu_{m,1}^{2}\right]  & =  \Ex{\| (\vec H_{m,1})^H\vec
    u_{m}\|_2^2} = \Ex{(\vec u_m)^H  \vec H_{m,1} (\vec H_{m,1})^H \vec u_m }
  = 1 \\ 
  \mathbb{E} \left[  \mu_{m,1}^{4}\right]  & \leq  \Ex{\left(\sum_{i=1}^{n_t}\|
    \vec h_{i}\|_2^2\right)^2} \leq n_r^2\\
\mathbb{E}^{\frac{1}{4}}\left[
  \mu_{m,1}^{8}\right]  & \leq  n_r,
\end{align*}
where $ \vec h_{i}$ is the $i$th column of  $\vec H_{m,1}$ and we used the
Cauchy-Schwarz inequality and the fact that $n_t \sum_{i=1}^{n_t}\| \vec h_{i}
\|_2^2 $ is chi-squared distributed with $n_tn_r$ degrees of freedom. 
\end{IEEEproof}
\begin{remark}
  The upper bound in Lemma \ref{lem:motivation} also holds (up to a
  constant) if we consider perfect link adaptation. Therefore, Corollary
  \ref{cor:scalingALG} is also true (up to a constant) if we consider
  perfect link adaptation.
\end{remark}
The following theorem shows that the scaling
$2^{-\frac{B}{2(n_{t}-1)}}$ can not be improved if we consider IA with
RVQ, as described in Section \ref{sec:Sched&FB}, and link adaptation.
\begin{theorem}
\label{theo:scalingIA} Under the assumptions of Theorem \ref{theo:scalingANY},
for sufficiently high SNR the average rate loss is bounded from below by
\begin{multline}
\mathbb{E}\left[  {\Delta r_{m}(\vec{\pi}_{H},\vec{\pi}_{V})}\right]  
\geq\max_{c_{1} > 0}\frac{1}{c_{1}}\left(  1-\frac{n_{t}^{2}\left(
n_{t}+1\right)  }{4\sqrt{c_{1}}\left(  n_{t}-1\right)  2^{-n_{t}}}\right) \\ 
\mathbb{E}\left[  \log\left(  1+\frac{c_{1}P\Ex{\mu_{m,b}^{2}}}{|\mathcal{S}%
_{b}|\left(  1+P\Ex{\mu_{m,b}^{2}}\right)  }\sqrt{\frac{4\cdot2^{-n_{t}}n_{t}%
-1}{n_{t}^{2}\left(  n_{t}+1\right)  }}2^{\frac{-B}{2\left(  n_{t}-1\right)
}}\right)  \right]  \\ 
-\log\left(  1+PK\sqrt{\mathbb{E}\left[  \mu_{m,b}^{4}\right]  }2^{\frac{-B}%
{n_{t}-1}}\right)
\end{multline}
where $|\mathcal{S}_{b}|$ satisfies the IA feasibility condition \eqref{eq:maxUE}, for all $b=1,\ldots,K$,
with equality. In particular, for some $c_{2}>0$
\begin{equation}
\mathbb{E}\left[  {\Delta r_{m}(\vec{\pi}_{H},\vec{\pi}_{V})}\right]  \geq
c_{2}\log\left(  1+\frac{P\mathbb{E}\left[  \mu_{m,b}^{2}\right]
}{|\mathcal{S}_{b}|\left(  1+P\mathbb{E}\left[  \mu_{m,b}^{2}\right]
\right)  }2^{\frac{-B}{2\left(  n_{t}-1\right)  }}\right)  +o\left(
2^{\frac{-B}{\left(  n_{t}-1\right)  }}\right)
\end{equation}
holds. 
\end{theorem}
The proof can be found in Appendix \ref{proof:scalingIA}.
Note that, the lower bound is bounded in $P$. Therefore it can not be
used for degrees of freedom analysis, where $P$ is taken to infinity.
But, Theorem \ref{theo:scalingIA} shows that the scaling
$2^{(-B/(2(n_t-1)))}$ can not be improved for finite SNR $P$.

\section{Simulations}

\subsection{Baseline}

As a baseline scheme we consider centralized IA which was proposed in
\cite{Schreck2011WSA} by the authors. The baseline scheme requires a
central processing unit which has global (quantized) CSI. Each user
$m$ quantizes and feeds back the channel matrix $\vec{H}_{m,l}$, for
all $l=1,\ldots,K$, to the central processing unit. The central
processing unit computes the transmit beamformers in an iterative
manner similar to Algorithm \ref{alg1} proposed in Section
\ref{sec:IA}. To quantize the channel matrices, we apply a scalar
quantization or a vector quantization scheme.
\subsubsection{Scalar Quantization}
Each user maps each element of the channel
matrix to an element of a scalar feedback codebook with $2^{B_{s}}$
elements. Scalar quantization (SQ) leads to a feedback load of $2Kn_{t}n_{r}B_{s}$
bits per user and per feedback message. As we will see, the feedback and
control overhead is significantly larger than for the proposed distributed
algorithm.

\subsubsection{Vector Quantization}
Vector quantization (VQ) is a popular quantization scheme for multi-antenna
channels. As a baseline we consider the following scheme which was
also used in \cite{Kim2012} and \cite{Krishnamachari2010}. Each user
$m$ quantizes the channel matrices $\vec H_{m,l}$, for all
$l=1,\ldots,K$, by applying RVQ (see Subsection \ref{sec:Sched&FB}) on the vector $\text{vec} (\vec
H_{m,l})$, where $\text{vec}(\vec X)$ stacks the columns of the matrix
$\vec X$  one over the other. 

\begin{figure}[ptb]
\centering
\includegraphics[width=.75\linewidth]{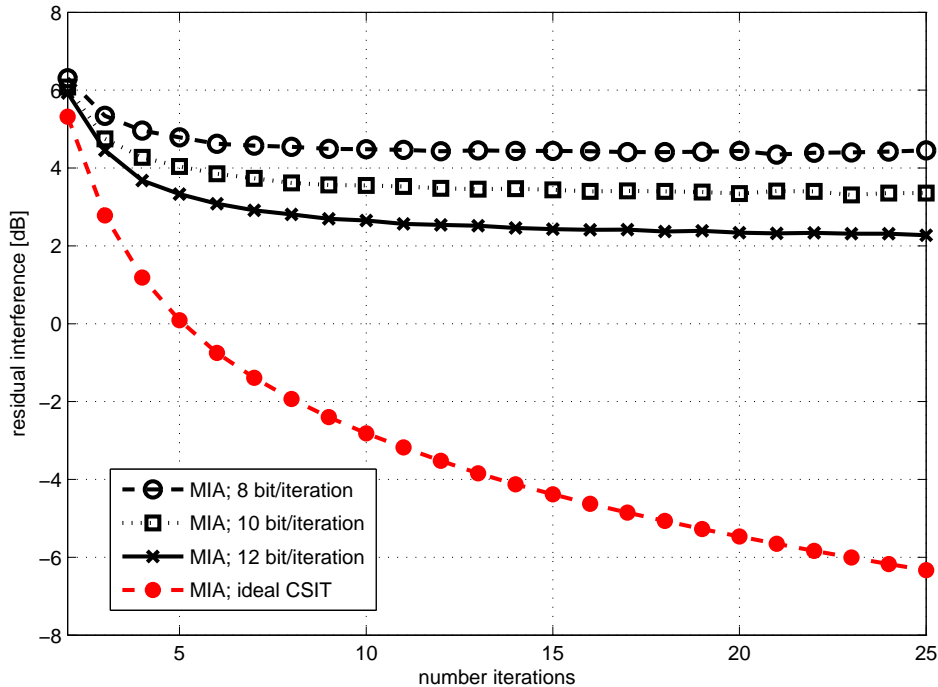}
%where an .eps filename suffix will be assumed under latex,
%and a .pdf suffix will be assumed for pdflatex
\caption{Spectral efficiency over SNR. Convergence of the proposed minimum interference algorithm
(Algorithm \ref{alg1}). Observation: The proposed algorithm converges
quickly.}%
\label{fig:conv}%
\end{figure}

\subsection{Simulation Setup}

\begin{figure}[ptb]
\centering
\includegraphics[width=.75\linewidth]{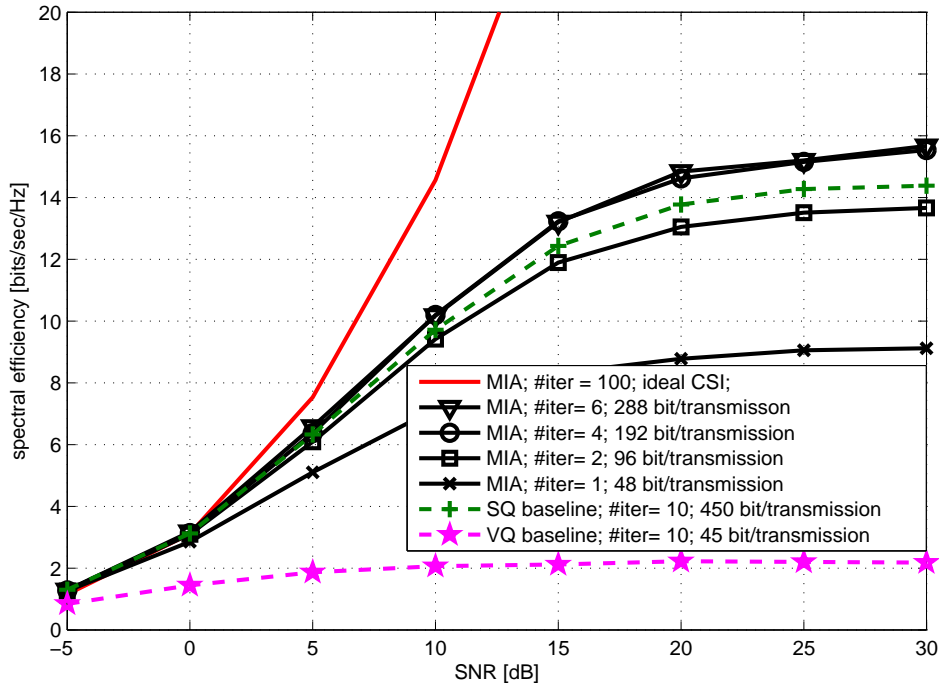}
%where an .eps filename suffix will be assumed under latex,
%and a .pdf suffix will be assumed for pdflatex
\caption{Spectral efficiency over SNR. Convergence of the proposed minimum interference algorithm
(Algorithm \ref{alg1}). The proposed algorithm converges quickly and
outperforms the base line with less than half the number of feedback bits and
iterations. }
\label{fig:sim1}
\end{figure}

In the simulations we consider a cellular network with $K=3$ base stations and
$U=9$ users. Each node is equipped with $n_{t}=n_{r}=5$ antennas. From
\cite{Schreck2012WSA} we know that IA is feasible if each base station $b$
serves $|\mathcal{S}_{b}|=3$ users. For the minimum interference algorithm (Algorithm \ref{alg1}) each
base station is assigned randomly to three users. Power allocation is assumed
to be uniform. The channels are modeled as described in Section
\ref{sec:sysmod}.

\subsection{Simulation Results}

First, we investigate the convergence of the proposed Algorithm
\ref{alg1}. Figure \ref{fig:conv} depicts the residual interference of
Algorithm \ref{alg1}. We observe that with quantized CSI the
residual interference converges rapidly to its minimum of approximately $4.5$
dB, $3.5$ dB and $2.5$ dB for $8$, $10$ and $12$ bit per user per iteration,
respectively. In contrast, with ideal CSI the residual interference keeps
decreasing with the number of iterations. We conclude that with
quantized CSI the number of iterations can be kept low ($\approx5$ iterations)
without loosing a significant part of the performance that can be achieved
with more iterations. Hence, the algorithm seems to be well suited for
practical applications where a small number of iterations is mandatory. 

This observation is further supported by Figure \ref{fig:sim1} which
depicts the spectral efficiency over the SNR for the minimum interference algorithm (Algorithm
\ref{alg1}). Each user uses an independent random codebook with
$2^{16}$ isotropic elements.  Again, we observe that the proposed IA
algorithm converges rapidly, i.e., going from $4$ to $6$ iterations
the performance is increased only slightly. This is a remarkable
results since a small number of iterations keeps the feedback load
small.

In addition, Figure \ref{fig:sim1} shows the performance of the SQ
baseline and VQ baseline schemes, defined above. Due to the
centralized approach the feedback load of these schemes does not
increase with the number of iterations. After 10 iterations the
baseline schemes have converged.  The SQ baseline is clearly
outperformed by the minimum interference algorithm with 4 iterations
and 192 bit feedback load per user and per transmission.  That is,
with the iterative minimum interference algorithm we require less than
half the number feedback bits to outperform the SQ baseline. The VQ
base line with a 15 bit random codebook (45 bit feedback per
transmission per user) performs very poorly. To obtain a better
performance significantly larger codebooks are required. However,
computing the feedback decision for larger codebooks becomes quickly
infeasible.

% \begin{figure}[ptb]
% \centering
% \includegraphics[width=.75\linewidth]{fig/cdf_MIA_MSA.eps}
% %where an .eps filename suffix will be assumed under latex,
% %and a .pdf suffix will be assumed for pdflatex
% \caption{CDF of per user spectral efficiency at SNR$=20$ dB. Comparing the MIA
%   (Algorithm \ref{alg1}) and the MSA (Algorithm \ref{alg2a}) with
%   quantized and ideal CSI. The MSA schedules less users but achieves a higher
%   spectral efficiency per scheduled user. }
% \label{fig:cdf}
% \end{figure}

% Figure \ref{fig:cdf} depicts the cdf of the per user spectral efficiency at
% SNR$=20$ dB. We
% observe that the MSA (Algorithm \ref{alg1}) schedules less users than the MIA
% (Algorithm \ref{alg2a}) but achieves a higher spectral efficiency for the
% scheduled users.  For both algorithms, with quantized CSI the mean spectral efficiency per user is
% approximately half of the  mean spectral efficiency that can be achieved with
% ideal CSI. Figure \ref{fig:sumcdf} depicts the CDF of the network spectral
% efficiency at SNR$=20$ dB. With quantized CSI the MSA outperforms the MIA. 

% \begin{figure}[ptb]
% \centering
% \includegraphics[width=.75\linewidth]{fig/sumcdf_MIA_MSA.eps}
% %where an .eps filename suffix will be assumed under latex,
% %and a .pdf suffix will be assumed for pdflatex
% \caption{CDF of network spectral efficiency at SNR$=20$ dB. Comparing the MIA
%   (Algorithm \ref{alg1}) and the MSA (Algorithm \ref{alg2a}) with
%   quantized and ideal CSI. The MSA schedules less users but achieves a higher
%   spectral efficiency per scheduled user. }
% \label{fig:sumcdf}
% \end{figure}

\section{Conclusion}

We introduced an improved metric for the performance evaluation of the
interfering broadcast channel. The improved metric captures the
throughput degradation due to quantized channel state information by
considering the beamformer offset and the link adaptation problem. We
obtained the relevant scaling laws and showed that they are
different from existing ones. Moreover, we provided an
iterative IA algorithm and corresponding feedback strategies which
achieve the derived scaling laws.

\bibliographystyle{IEEEtran}
\bibliography{ref}

\appendices
\section{Proof of Theorem \ref{theo:scalingANY}}

\label{proof:scalingANY} 

Before proving Theorem \ref{theo:scalingANY} we prove two lemmas which are
required in the proof.

% For RVQ where $\mathcal{V}=\left(  \vec{v}%
% _{1},...,\vec{v}_{2^{B}}\right)  \subset\mathbb{S}^{n_t-1},\vec{h}_{k,l}%
% \in\mathbb{S}^{n_t-1},\forall k,l,$ are independent random variables, the
% quantization error $Z=\underset{\vec{v}\in\mathcal{V}}{\min}(1-|\langle
% \vec{h}_{k,l},\vec{v}\rangle|^{2})$ is the minimum of $2^{B}$ beta$\left(
% n_{t}-1,1\right)  $ distributed random variables (see e.g. \cite[Theorem
% 2]{Jindal2006}). This is true even when $Z$ is conditioned on $\vec{h}_{k,l}$.
% The following results on the moments of $Z$ hold.
\begin{lemma}
\label{lem:q_error} 
Let $\vec x \in \mathbb{S}^{n_t-1}$ be an independent isotropic random
vector and $\mathcal{V}=\left( \vec{v}_{1},...,\vec{v}_{2^{B}}\right)
\subset\mathbb{S}^{n_t-1}$ be a collection of $2^B$ independent
isotropic random vectors. If we define
\begin{equation}
Z:=\underset{\vec{v}\in\mathcal{V}}{\min}(1-|\langle \vec
x,\vec{v}\rangle|^{2}),\end{equation}
then for some $n\geq1$ we have
\begin{equation}
\left(  \frac{n_{t}-1}{n_{t}}2^{\frac{-B}{n_{t}-1}}\right)  ^{n}\leq
\mathbb{E}\left[  {Z^{n}}\right]  \leq\left(  2^{\frac{-B}{n_{t}-1}}\right)
^{n}\leq\left(  \frac{n_{t}}{n_{t}-1}\mathbb{E}\left[  {Z}\right]  \right)
^{n}.
\end{equation}
\end{lemma}
\begin{IEEEproof}
  Since, $1-|\langle \vec x,\vec{v}\rangle|^{2}$ is beta distributed
  with parameters $n_{t}-1$ and $1$ \cite{Jindal2006}, $Z$ is the minimum of $2^{B}$
  $\text{beta}( n_{t}-1,1)$ distributed random variables. 
  Therefore,
  $\mathbb{E}\left[ {Z}\right] \geq\frac{n_{t}-1}{n_{t}}2^{\frac
    {-B}{n_{t}-1}}$ (see e.g. \cite{Jindal2006}) and by Jensen's
  inequality we get the lower bound
  \begin{equation}
  \mathbb{E}\left[  {Z^{n}}\right]  \geq\mathbb{E}\left[  {Z}\right]  ^{n}%
   = \left(  \frac{n_{t}-1}{n_{t}}2^{\frac{-B}{n_{t}-1}}\right)  ^{n}.
  \end{equation}
  For the upper bound we use \cite[Lemma 1]{Au-Yeung2007} which states
  that $\Pr\left( Z >x\right) =\left(1-x^{n_{t}-1}\right)^{2^{B}}$.
  Thus, $\Pr\left( Z^{n}>x\right) = \Pr\left( Z > x^{1/n}\right) = \left(
    1-x^{\frac{n_{t}-1}{n}}\right)^{2^{B}}$, with $n\geq1$, and
  therefore
  \begin{equation}
  \mathbb{E}\left[ {Z^{n}}\right] \leq\left(
    2^{\frac{-B}{n_{t}-1}}\right) ^{n}.
  \end{equation}
  Using,  $\mathbb{E}\left[ {Z}\right] \geq\frac{n_{t}-1}{n_{t}}2^{\frac
    {-B}{n_{t}-1}}$ once more the second upper bound follows. 
\end{IEEEproof}
Let us now define the following metric.
\begin{mydef}
  \begin{equation}
    \omega\left(  \vec{x},\vec{y}\right)  :=\max_{\vec{w}\in
      \mathbb{S}^{n_{t}-1}}\left\vert |\langle\vec{x},\vec{w}\rangle
      |^{2}-|\langle\vec{y},\vec{w}\rangle|^{2}\right\vert
\label{eq:new_metric}%
\end{equation}
\end{mydef}
As we will see, this metric essentially dictates the rate loss gap
in Theorem \ref{theo:scalingANY}. We have the following lemma, which
shows that this metric is equal to the chordal distance. 
\begin{lemma}
\label{lem:ExChor} Let $\vec{x}\in\mathbb{S}^{n_t-1}$ and $\vec{y}%
\in\mathbb{S}^{n_t-1}$ be unit norm vectors, then
\begin{equation}
\max_{\vec{w}\in\mathbb{S}^{n_t-1}}\left\vert |\langle\vec{x},\vec{w}%
\rangle|^{2}-|\langle\vec{y},\vec{w}\rangle|^{2}\right\vert =\sqrt
{1-|\langle\vec{x},\vec{y}\rangle|^{2}}.\label{eq:disMD}%
\end{equation}
\end{lemma}
\begin{IEEEproof}
We have
\begin{equation}
\left\vert |\langle\vec{x},\vec{w}\rangle|^{2}-|\langle\vec{y},\vec{w}%
\rangle|^{2}\right\vert =\left\vert \vec{w}^{H}(\vec{x}\vec{x}^{H}-\vec{y}%
\vec{y}^{H})\vec{w}\right\vert.
\end{equation}
Consider the matrix $\vec{A}:=\vec{x}\vec{x}^{H}-\vec{y}\vec{y}^{H}$. Since, $\rank{\cdot}$ is a subadditive function we have that the matrix
$\vec{A}$ has maximum rank of two and, therefore, has only two non-zero eigenvalues
$\lambda_{1}$ and $\lambda_{2}$. But the matrix $\vec{A}$ is trace-less as
well
\begin{equation}
\text{Tr}(\vec{A})=\lambda_{1}+\lambda_{2}=\text{Tr}(\vec{x}\vec{x}^{H}%
-\vec{y}\vec{y}^{H})=\lVert\vec{x}\rVert_{2}^{2}-\lVert\vec{y}\rVert_{2}%
^{2}=0. \label{eq:lambda=0}
\end{equation}
Therefore, $\lambda_{1}=-\lambda_{2}$ must hold. On the other hand, we get
from the Frobenius norm $\lVert\vec{A}\rVert_{F}^{2}=\text{Tr}(\vec{A}%
^{H}\vec{A})$ that
\begin{equation}%
\begin{split}
\text{Tr}(\vec{A}^{H}\vec{A})  &  =\lambda_{1}^{2}+\lambda_{2}^{2}=\lVert
\vec{x}\rVert_{2}^{4}+\lVert\vec{y}\rVert_{2}^{4}-2|\langle\vec{x},\vec
{y}\rangle|^{2}\\
&  =2(1-|\langle\vec{x},\vec{y}\rangle|^{2})
\end{split}
\end{equation}
Thus, using $\lambda_{1}=-\lambda_{2}$ we get for the two non-zero
eigenvalues
\begin{equation}
|\lambda_{1}|=|\lambda_{2}|=\sqrt{1-|\langle\vec{x},\vec{y}\rangle|^{2}},
\label{eq:lambda}%
\end{equation}
which proves the claim.
\end{IEEEproof}
We are now ready to prove Theorem \ref{theo:scalingANY}.
\begin{IEEEproof}
Fix the user selection $\mathcal{S}$ and the transmit beamformers
$\vec{\pi}$. Denote the optimal receive filter as $\vec{u}_{m}^{\ast}$ with
respect to the collection $H$, we have $\vec{\hat{h}}_{m,b}=(\vec{H}%
_{m,b})^{H}\vec{u}_{m}^{\ast}$, $\mu_{m,b}=\Vert\vec{\hat{h}}_{m,b}%
\Vert_{2}$ and $\vec{h}_{m,b}=\vec{\hat{h}}_{m,b}/\mu_{m,b}$ for all
$b=1,...,K$. Hence, the achievable rate of user $m\in\mathcal{S}_{b}$ is
\begin{equation}
r_{m}(\vec{\pi},\mathcal{S};H)=\log\left(  1+\frac{\frac{P\mu_{m,b}^{2}%
}{|\mathcal{S}_{b}|}|\langle\vec{h}_{m,b},\vec{\pi}(m)\rangle|^{2}}%
{1+\sum_{l=1}^{K}\sum_{\multi{k\in\mathcal{S}_{l}}{k\neq m}
}\frac{P\mu_{m,l}^{2}}{|\mathcal{S}_{l}|}|\langle\vec{h}_{m,l},\vec{\pi
}(k)\rangle|^{2}}\right).
\end{equation}
Since, the base station does not know the channels $H$ it must use the imperfect CSI $V=\{\vec
{\hat{v}}_{m,b}=\mu_{m,b}\vec{v}_{m,b}:m\in\mathcal{U}
,b=1,\ldots,K\}$. 
Based on $V$ the base station estimates the rates achievable by user $m$ as 
\begin{equation}
r_{m}(\vec{\pi},\mathcal{S};V)=\log\left(  1+\frac{\frac{P\mu_{m,b}^{2}%
}{|\mathcal{S}_{b}|}|\langle\vec{v}_{m,b},\vec{\pi}(m)\rangle|^{2}}%
{1+\sum_{l=1}^{K}\sum_{\multi{k\in\mathcal{S}_{l}}{k\neq m}%
}\frac{P\mu_{m,l}^{2}}{|\mathcal{S}_{l}|}|\langle\vec{v}_{m,l},\vec{\pi
}(k)\rangle|^{2}}\right)  .
\end{equation}
For the ease of presentation, we define the following variables (index
$m$ omitted), 
\begin{align}
\Phi_{b,k}  &  :=\frac{P\mu_{m,b}^{2}}{|\mathcal{S}_{b}|}|\langle\vec{h}
_{m,b},\vec{\pi}(k)\rangle|^{2} \label{eq:effH}\\
\Psi_{b,k}  &  :=\frac{P\mu_{m,b}^{2}}{|\mathcal{S}_{b}|}|\langle\vec{v}
_{m,b},\vec{\pi}(k)\rangle|^{2}\label{eq:effV}\\
\Delta_{b,k}  &  :=\Phi_{b,k}-\Psi_{b,k}\nonumber\\
\Psi_{\Sigma} &  = \left(1+\sum_{l=1}^{K}\sum_{k\in\mathcal{S}_{l}}\Psi_{l,k}\right)^{-1} \nonumber\\
\Phi_{\Sigma}  &  =\left(1+\sum_{l=1}^{K}\sum_{
\multi{k\in\mathcal{S}_{l}}{k\neq m}
}\Phi_{l,k}\right)^{-1}.\nonumber
\end{align}
Equation \eqref{eq:effH} and  \eqref{eq:effV} can be interpreted
as the effective received SNR of user $m$ through the channels $\vec
h_{m,b}$ and $\vec v_{m,b}$, respectively. 
Further, define $\delta := |r_{m}(\vec{\pi},\mathcal{S};H) -
r_{m}(\vec{\pi},\mathcal{S};V)|$, which can bounded from above by
\begin{align}
\delta \leq &  \max_{\mathcal{S}\subseteq U}\max_{\vec{\pi}:\set S \rightarrow \Sn}|r_{m}(\vec{\pi},\mathcal{S}%
;H)-r_{m}(\vec{\pi},\mathcal{S};V)|\nonumber\\
&  =\max_{\mathcal{S}\subseteq U}\max_{\vec{\pi}:\set S \rightarrow \Sn}\left\vert \log\left(  \frac
{1+\sum_{l=1}^{K}\sum_{k\in\mathcal{S}_{l}}\Phi_{l,k}}{1+\sum_{l=1}^{K}%
\sum_{k\in\mathcal{S}_{l}}\Psi_{l,k}}\right)  +\log\left(  \frac{1+\sum
_{l=1}^{K}\sum_{%
\multi{k\in\mathcal{S}_{l}}{k\neq m}%
}\Psi_{l,k}}{1+\sum_{l=1}^{K}\sum_{%
\multi{k\in\mathcal{S}_{l}}{k\neq m}%
}\Phi_{l,k}}\right)  \right\vert \nonumber\\
&  =\max_{\mathcal{S}\subseteq U}\max_{\vec{\pi}:\set S \rightarrow \Sn}\left\vert \log\left(  1+\sum
_{l=1}^{K}\sum_{k\in\mathcal{S}_{l}}\Psi_{\Sigma}\Delta_{l,k}\right)
+\log\left(  1+\sum_{l=1}^{K}\sum_{%
\multi{k\in\mathcal{S}_{l}}{k\neq m}%
}\Phi_{\Sigma}\left(  -\Delta_{l,k}\right)  \right)  \right\vert. \label{eq:cor1Tight}
\end{align}

Using $\log\left(  1+a\right)  +\log\left(  1+b\right)  \leq2\log\left(
1+\frac{1}{2}(a+b)\right)$, with $a,b>-1$, yields
\begin{align*}
\delta &  \leq\max_{\mathcal{S}\subseteq U}\max_{\vec{\pi}:\set S \rightarrow \Sn}\left\vert \log\left(
1+\sum_{l=1}^{K}\sum_{k\in\mathcal{S}_{l}}\Psi_{\Sigma}\cdot\Delta
_{l,k}\right)  \right. \\ & \left. \quad + \log\left(  1+\sum_{l=1}^{K}\sum_{k\in\mathcal{S}_{l}}%
\Phi_{\Sigma}\cdot\left(  -\Delta_{l,k}\right)  +\sum_{l=1}^{K}\Phi_{\Sigma
}\cdot\Delta_{l,m}\right)  \right\vert \\
&  =2\, \max_{\mathcal{S}\subseteq U}\max_{\vec{\pi}:\set S \rightarrow \Sn} \left\vert \log\left(  1+\frac
{1}{2}\left(  \Psi_{\Sigma}-\Phi_{\Sigma}\right)  \sum_{l=1}^{K}\sum
_{k\in\mathcal{S}_{l}}\Delta_{l,k}+\frac{1}{2}\sum_{l=1}^{K}\Phi_{\Sigma}%
\cdot\Delta_{l,m}\right)  \right\vert \\
&  \leq2\, \max_{\mathcal{S}\subseteq U}\max_{\vec{\pi}:\set S \rightarrow
  \Sn} \log\left(  1+\frac{1}%
{2}\left\vert \Psi_{\Sigma}-\Phi_{\Sigma}\right\vert \left\vert \sum_{l=1}%
^{K}\sum_{k\in\mathcal{S}_{l}}\Delta_{l,k}\right\vert +\frac{1}{2}\sum
_{l=1}^{K}\left\vert \Delta_{l,m}\right\vert \right).
\end{align*}
Since,
\begin{align*}
\Psi_{\Sigma}-\Phi_{\Sigma}  &  =\frac{1}{\Psi_{\Sigma}^{-1}\Phi_{\Sigma}%
^{-1}}\left( \sum_{l=1}^{K}\sum_{\multi{k\in\mathcal{S}_{l}}{k\neq m}%
}\Phi_{l,k}-\sum_{l=1}^{K}\sum_{k\in\mathcal{S}_{l}}\Psi_{l,k}\right)  \leq\frac{1}{\Psi_{\Sigma}^{-1}\Phi_{\Sigma}^{-1}}\sum_{l=1}^{K}\sum
_{k\in\mathcal{S}_{l}} | \Delta_{l,k}|
\end{align*}
holds, we have the result
\begin{align*}
\delta &  \leq2 \, \max_{\mathcal{S}\subseteq U} \max_{\vec \pi : \set S  \rightarrow \Sn}\log\left(  1+\frac{1}%
{2}\left(  \sum_{l=1}^{K}\sum_{k\in\mathcal{S}_{l}}\left\vert \Delta
_{l,k}\right\vert \right)  ^{2}+\frac{1}{2}\sum_{l=1}^{K}\left\vert
\Delta_{l,m}\right\vert \right) \\
& \leq2\log\left(  1+\frac{1}{2}\max_{\mathcal{S}\subseteq U}\sum_{l_{1}=1}^K\sum_{k_{1}\in\mathcal{S}_{l_{1}}}\sum_{l_{2}=1}^K\sum_{k_{2}\in\mathcal{S}_{l_{2}}}\max_{\vec \pi : \set S  \rightarrow \Sn }\left\vert \Delta_{l_{1},k_{1}
}\right\vert \max_{\vec \pi : \set S  \rightarrow \Sn }\left\vert \Delta_{l_{2},k_{2}}\right\vert
\right. \\ & \quad \left. +\frac{1}{2}\sum_{l=1}^{K}\max_{\vec \pi : \set S  \rightarrow \Sn}\left\vert \Delta_{l,m}\right\vert
\right).
\end{align*}
Now, observe that
\begin{align*}
\max_{\vec \pi : \set S  \rightarrow \Sn}\left\vert \Delta_{l,k}\right\vert  &  =\frac{P\mu_{m,l}^{2}%
}{|\mathcal{S}_{l}|}\max_{\vec \pi : \set S  \rightarrow \Sn}\left\vert |\langle\vec{h}_{m,l},\vec{\pi
}(k)\rangle|^{2}-|\langle\vec{v}_{m,l},\vec{\pi}(k)\rangle|^{2}\right\vert \\
&  =\frac{P\mu_{m,l}^{2}}{|\mathcal{S}_{l}|}\max_{\vec{x}\in\mathbb{S}%
^{n_{t}-1}}\left\vert |\langle\vec{h}_{m,l},\vec{x}\rangle|^{2}-|\langle
\vec{v}_{m,l},\vec{x}\rangle|^{2}\right\vert \\
&  =\frac{P\mu_{m,l}^{2}}{|\mathcal{S}_{l}|}\omega\left(  \vec{h}_{m,l}%
,\vec{v}_{m,l}\right),
\end{align*}
where $\omega(\cdot,\cdot)$ was defined in \eqref{eq:new_metric} and
the last term $\frac{P\mu_{m,b}^{2}}{|\mathcal{S}_{b}|}\omega\left(
  \vec{h}_{m,l},\vec{v}_{m,l}\right) $ is actually independent of
$k$. Taking expectation and applying Jensen's inequality we obtain
\begin{multline}
\delta \leq2\log\Biggl(  1+\frac{P}{2}\sum_{l_{1}=1}^{K}\sum_{l_{2}=1}%
^{K}\mathbb{E}\left[  \mu_{m,l_{1}}^{2}\omega\left(  \vec{h}_{m,l_{1}},\vec
{v}_{m,l_{1}}\right)  \mu_{m,l_{2}}^{2}\omega\left(  \vec{h}_{m,l_{2}},\vec
{v}_{m,l_{2}}\right)  \right]  \\ +\frac{P}{2}\sum_{l=1}^{K}\mathbb{E}\left[
\mu_{m,l}^{2}\omega\left(  \vec{h}_{m,l},\vec{v}_{m,l}\right)  \right]
\Biggr). \label{eq:step1}
\end{multline}
Now, for $l_{1}\neq l_{2}$ we have from the Cauchy-Schwarz inequality
\begin{multline}
  \mathbb{E}\left[  \mu_{m,l_{1}}^{2}\omega\left(  \vec{h}_{m,l_{1}},\vec
{v}_{m,l_{1}}\right)  \mu_{m,l_{2}}^{2}\omega\left(  \vec{h}_{m,l_{2}},\vec
{v}_{m,l_{2}}\right)  \right] =\mathbb{E}\left[  \mu_{m,l_{1}}^{2}\omega\left(  \vec{h}_{m,l_{1}},\vec
{v}_{m,l_{1}}\right)  \right]  \mathbb{E}\left[  \mu_{m,l_{2}}^{2}%
\omega\left(  \vec{h}_{m,l_{2}},\vec{v}_{m,l_{2}}\right)  \right] \\
  \leq\mathbb{E}^{\frac{1}{2}}\left[  \mu_{m,l_{1}}^{4}\right]
\mathbb{E}^{\frac{1}{2}}\left[  \omega^{2}\left(  \vec{h}_{m,l_{1}},\vec
{v}_{m,l_{1}}\right)  \right]  \mathbb{E}^{\frac{1}{2}}\left[  \mu_{m,l_{2}%
}^{4}\right]  \mathbb{E}^{\frac{1}{2}}\left[  \omega^{2}\left(  \vec
{h}_{m,l_{2}},\vec{v}_{m,l_{2}}\right)  \right] \label{eq:step2}
\end{multline}
and for $l_{1}=l_{2}$ we have  from the Cauchy-Schwarz inequality
\begin{align}
  \mathbb{E}\left[  \mu_{m,l_{1}}^{4}\omega^{2}\left(  \vec{h}_{m,l_{1}}%
,\vec{v}_{m,l_{1}}\right)  \right] 
&  \leq\mathbb{E}\left[  \mu_{m,l_{1}}^{4}\omega\left(  \vec{h}_{m,l_{1}}%
,\vec{v}_{m,l_{1}}\right)  \right] \nonumber \\
&  \leq\mathbb{E}^{\frac{1}{4}}\left[  \mu_{m,l_{1}}^{8}\right]
\mathbb{E}^{\frac{1}{2}}\left[  \omega^{2}\left(  \vec{h}_{m,l_{1}},\vec
{v}_{m,l_{1}}\right)  \right]. \label{eq:step3}
\end{align}
Using Lemma \ref{lem:ExChor} and
Lemma \ref{lem:q_error} we get
\begin{align}
\mathbb{E}\left[  \omega^{2}\left(  \vec{h}_{m,l},\vec{v}_{m,l}\right)
\right]   &  =\mathbb{E}\left[  \underset{\vec{v}\in\mathcal{V}}{\min
} \, \left( 1-|\langle\vec{h}_{k,l},\vec{v}\rangle|^{2} \right)\right]   <2^{\frac{-B}{n_{t}-1}}. \label{eq:step4}
\end{align} 
Such that, the claim follows by plugging \eqref{eq:step2},
\eqref{eq:step3} and \eqref{eq:step4} in  \eqref{eq:step1}. 
% \begin{equation}
% \delta \leq2\log\left(  1+\frac{P}{2}\left[  K\left(  K-1\right)  \mathbb{E}%
% \left[  \mu_{m,1}^{4}\right]  2^{\frac{-B}{n_{t}-1}}+2^{\frac{-B}{2(n_{t}-1)}%
% }\left(  \mathbb{E}^{\frac{1}{4}}\left[  \mu_{m,1}^{8}\right]  \mathbb{+}%
% K\mathbb{E}^{\frac{1}{2}}\left[  \mu_{m,1}^{2}\right]  \right)  \right]
% \right).
% \end{equation}
\end{IEEEproof}

\section{Proof of Theorem \ref{theo:scalingIA}}

\label{proof:scalingIA} 

The following lemma allows us to bound the expected value of certain
concave functions from below. The lemma is a partial reverse of
Jensen's inequality when certain conditions on the moments are
fulfilled; more precisely, if $\left( 1-\sqrt{\frac{\mathbb{E}\left[
        {z^{2}}\right] }{c_{1}\mathbb{E}\left[ {z}\right]
      ^{2}}}\right) \geq c_3$ holds, with $c_1>0$ and $c_3\neq 0$
being constants.  Note that this is exactly the case for the
quantization error, as we will see in the proof of Theorem
\ref{theo:scalingIA}. 

The following lemma is a partial reverse of Jensen's inequality for
super linear functions. The lemma will be useful since $\log(1+x)$ is
a super linear function.  
\begin{lemma}
  \label{lem:lowerBound} If $f:\mathbb{R}\rightarrow\mathbb{R}$ is
  superlinear, then for any constant $c_{1}>0$ and any random variable
  $z>0$,
\begin{equation}
\mathbb{E}\left[  {f(z)}\right]  \geq\frac{f(c_{1}\mathbb{E}\left[
{z}\right]  )}{c_{1}}\left(  1-\sqrt{\frac{\mathbb{E}\left[  {z^{2}}\right]
}{c_{1}\mathbb{E}\left[  {z}\right]  ^{2}}}\right)
\end{equation}
holds.
\end{lemma}
\begin{IEEEproof}
  By assumption $f(z)$ is superlinear and therefore $f(0)=0$. Thus,
  for any $x$ and $y$ and any $t\in [0,1]$, $f(tx + (1-t)y) \geq tf(x)
  + (1-t)f(y)$. Setting $x=0$ and $y=z^\ast$,
  $f((1-t)z^\ast) \geq \frac{z^\ast}{z^\ast}(1-t)f(z^\ast)$.  Hence,
  for any $0\leq z \leq z^{\ast}$,
  \begin{equation}
    \label{eq:concave}
    \frac{f(z)}{z} \geq \frac{f(z^\ast)}{z^\ast} 
  \end{equation}
  For any $z^{\ast}>0$ we have
  \begin{align}
    \mathbb{E}\left[  {f(z)}\right]   &  \geq\mathbb{E}\left[  {f(z)\mathbb{I}%
        \left\{  z\leq z^{\ast}\right\}  }\right]  \nonumber\\
    &  =\mathbb{E}\left[  {\frac{f(z)}{z}z\mathbb{I}\left\{  z\leq z^{\ast
          }\right\}  }\right]  \nonumber\\
    &  \geq\frac{f(z^{\ast})}{z^{\ast}}\mathbb{E}\left[  {z\mathbb{I}\left\{
          z\leq z^{\ast}\right\}  }\right]  \label{eq:conc}\\
    &  =\frac{f(z^{\ast})}{z^{\ast}}\mathbb{E}\left[  {z(1-\mathbb{I}\left\{
          z\ >z^{\ast}\right\}  )}\right]  \nonumber\\
    &  =\frac{f(z^{\ast})}{z^{\ast}}\left(  \mathbb{E}\left[  {z}\right]
      -\mathbb{E}\left[  {z\mathbb{I}\left\{  z\ >z^{\ast}\right\}  }\right]
    \right)  \\
    &  \geq \frac{f(z^{\ast})}{z^{\ast}}\left(  \mathbb{E}\left[  {z}\right]
      -\sqrt{\mathbb{E}\left[  {z^{2}}\right]  }\sqrt{\mathbb{E}\left[
          {\mathbb{I}\left\{  z\ >z^{\ast}\right\}  }\right]  }\right)  \label{eq:CS}\\
    &  =\frac{f(z^{\ast})}{z^{\ast}}\mathbb{E}\left[  {z}\right]  \left(
      1-\sqrt{\frac{\mathbb{E}\left[  {z^{2}}\right]  }{\mathbb{E}\left[
            {z}\right]  ^{2}}}\sqrt{\Pr(z\ >z^{\ast})}\right)  \nonumber\\
    &  \geq \frac{f(z^{\ast})}{z^{\ast}}\mathbb{E}\left[  {z}\right]  \left(
      1-\sqrt{\frac{\mathbb{E}\left[  {z^{2}}\right]  }{\mathbb{E}\left[
            {z}\right]  z^{\ast}}}\right). \label{eq:Mark}%
  \end{align}
Inequality \eqref{eq:conc} follows from \eqref{eq:concave}, \eqref{eq:CS} follows from the Cauchy-Schwarz inequality and
\eqref{eq:Mark} follows from Markov's inequality. If we choose $z^{\ast}%
=c_{1}\mathbb{E}\left[  {z}\right]  $ the claim follows.
% \begin{figure}[ptb]
% \centering
% \includegraphics[width=.5\linewidth]{fig/lemma4.eps}
% \caption{Example of a function satisfying Lemma \ref{lem:lowerBound}. For the Concave function $f(z)$ we have $f(z) \geq \frac{f(z^*)}{z^*}z$, for all $0\leq z \leq z^*$.}
% \label{fig:lemma4}
% \end{figure}
% \begin{equation}
% \mathbb{E}\left[  {f(z)}\right]  \geq\frac{f(c_{1}\mathbb{E}\left[
% {z}\right]  )}{c_{1}}\left(  1-\sqrt{\frac{\mathbb{E}\left[  {z^{2}}\right]
% }{c_{1}\mathbb{E}\left[  {z}\right]  ^{2}}}\right)  .
% \end{equation}
\end{IEEEproof}

% \footnote{We like to thank an anonymous reviewer for
  % pointing this out.}
The following lemma is a modification of Lemma
\ref{lem:ExChor}.

\begin{lemma}
\label{lem:ExChor2}

Let $\vec{x}\in\mathbb{S}^{n_t-1}$ and $\vec{y}\in\mathbb{S}^{n_t-1}$
be unit norm vectors. If $\vec{w}$ is uniformly distributed on the
unit sphere $\mathbb{S}^{n_t-1}$ and $m\geq2$, then
\begin{equation}
\frac{4\cdot2^{-n_t}}{n_t\left( n_t+1\right) }\left( 1-|\langle\vec{x},\vec{y}\rangle
  |^{2}\right) ^{\frac{m}{2}}
\leq
\mathbb{E}\left[ {\max\left\{|\langle\vec{x},\vec{w}\rangle|^{2}-|\langle\vec{y},\vec{w}\rangle|^{2},0\right\}
  }^{m}\right] 
\leq
\left(1-|\langle\vec{x},\vec{y}\rangle|^{2}\right) ^{\frac{m}{2}}
\end{equation} 
holds.
\end{lemma}
\begin{IEEEproof}
We have
\begin{equation}
\left\vert |\langle\vec{x},\vec{w}\rangle|^{2}-|\langle\vec{y},\vec{w}%
\rangle|^{2}\right\vert =\left\vert \vec{w}^{H}(\vec{x}\vec{x}^{H}-\vec{y}%
\vec{y}^{H})\vec{w}\right\vert.
\end{equation}
Consider the eigen-decomposition of the Hermitian matrix
$\vec{A}:=\vec{x}
\vec{x}^{H}-\vec{y}\vec{y}^{H}=\vec{Q}\vec{\Lambda}\vec{Q}^{H}$.
Since, $\vec A$ has maximum rank of $2$, it has at most two non-zero
eigenvalues. Therefore, the diagonal matrix $\vec{\Lambda}$ can be
written as $\vec{\Lambda} = \diag{\lambda_1,\lambda_2,0,\ldots,0}$, with
$\lambda_1\leq \lambda_2$. Since, $\vec Q$ is a Hermitian matrix,
with columns given by the eigenvectors of $\vec{A}$, and ${\vec{w}}$
is uniformly distributed on $\mathbb{S}^{n_t-1}$, the following is
true 
\begin{equation}
\mathbb{E}\left[  \left(  {\vec{w}^{H}\vec{A}\vec{w}}\right)  ^{m}\right]
=\mathbb{E}\left[  \left(  {\vec{w}^{H}\vec{\Lambda}\vec{w}}\right)
^{m}\right]  .
\end{equation}
According to \eqref{eq:lambda=0} $\lambda_{1}=-\lambda_{2}$. Thus, we have
\begin{align}
\mathbb{E}\left[  \max\left\{  \vec{w}^{H}\vec{A}\vec{w},0\right\}
^{m}\right]   &  =\mathbb{E}\left[  \max\left\{  {\lambda_{1}|w_{1}|}%
^{2}-{\lambda_{1}|w_{2}|}^{2},0\right\}  ^{m}\right]  \nonumber\\
&  \geq\lambda_{1}^{m} \mathbb{E}\left[  \max\left\{  2{|w_{1}|}^{2}-1,0\right\}  ^{m}\right]
 \label{eq:in1} \\
%&   \geq \lambda_{1}^{n_t}  \mathbb{E}\left[  \max\left\{  2{|w_{1}|}^{2}-1,0\right\}  ^{2}\right] % ^{\frac{n_t}{2}} \label{eq:in2} \\
& = \lambda_{1}^{m}  \mathbb{E}\Bigl[   (2{|w_{1}|}^{2}-1)^{m} \,
\bigl| \,
  |w_1|^2 \geq 1/2  \Bigr] \label{eq:pos}\\ 
& = \lambda_{1}^{m} \int_{\sqrt{1/2}}^{1}(2\varepsilon^{2}-1)^{m}
\, d\mu(\varepsilon). \label{eq:int}
% \\
% & = \lambda_{1}^{m} \int_{\sqrt{1/2}}^{1}(2\varepsilon^{2}-1)^{m}
% \mu'(\varepsilon) \, d \varepsilon,
\end{align}
Inequality \eqref{eq:in1} holds since $\|\vec w \|_2 = 1$ and therefore
${|w_{1}|^{2}+|w_{2}|^{2}\leq1}$.  
% Inequality \eqref{eq:in2} follows from Jensen's inequality.  
Equation \eqref{eq:pos} is true because $2{|w_{1}|}^{2}-1$ is
non-negative only for $|w_{1}|^2\geq 1/2$.  
In \eqref{eq:int}
$\mu(\varepsilon)$ is the Haar-measure of $\{ \vec w \in \Sn : |w_1|\leq
\varepsilon  \}$.  
Now we apply a result by Rudin
\cite[page 15 equation (2)]{Rudin1980} for functions on the sphere
$\mathbb{S}^{n_t-1}$ in one parameter. We have, for some function
$f:\Reals \rightarrow \Reals$ and a normalized measure $\sigma(\Sn)=1$,
\begin{align}
  \int_{\Sn} f(\sigma) \, d\sigma & =\frac{n_t-1}{\pi} \int_0^{2\pi} d\Theta
  \int_0^1 (1-r^2)^{n_t-2} f(r) r \, dr\nonumber\\
& = 2(n_t-1) \int_0^1 (1-r^2)^{n_t-2} f(r) r \, dr. \label{eq:rudin}
\end{align}
By setting $f(r) := \lambda_1^{m}(2r^2 - 1)^{m}
\chi_{\left[\sqrt{1/2} ,1\right]}(r) $, where $\chi_{I}(x)$ is the
characteristic function, the lower bound is proved as follows. 
Plugging $f(r)$ in \eqref{eq:rudin} and using \eqref{eq:int} we have
\begin{align}
  \mathbb{E}\left[ \max\left\{ \vec{w}^{H}\vec{A}\vec{w},0\right\}
    ^{m}\right] \geq & \lambda_{1}^{m} \int_{\sqrt{1/2}}^1 (2r^2
  - 1)^{m} \underbrace{2(n_t-1)(1-r^2)^{n_t-2} r \,
    dr}_{d \mu(r)} \nonumber\\
  &= \lambda_{1}^{m} (n_t-1) \int_{1/2}^1 (2u - 1)^{m}
  (1-u)^{n_t-2}  \, du \label{eq:u=r2} \\
  &= \lambda_{1}^{m}(n_t-1) \int_{1/2}^1 (2(u - 1)+1)^{m}
  (1-u)^{n_t-2}  \, du \\
  &= \lambda_{1}^{m}(n_t-1) \int_0^{1/2} (1 - 2v)^{m}
  v^{n_t-2}  \, dv  \label{eq:v=1-u}\\
  & = 4\lambda_{1}^{m} \frac{ 2^{-n_t}
  }{n_t(n_t+1)}. \label{eq:intSolve}
\end{align}
Equation \eqref{eq:u=r2} is obtained by substituting $u =r^2$ and in
\eqref{eq:v=1-u} we substituted $v = 1-u$. Finally, \eqref{eq:intSolve} follows by
solving the integral. Using  \eqref{eq:lambda} in the proof of Lemma \ref{lem:ExChor}, which states that the largest eigenvalue of $A$ is
$\lambda_1 = \sqrt{1-|\langle \vec x, \vec y \rangle|^2}$, the lower
bound is obtained. 

The upper bound follows,
 since $0 \leq \max\left\{  {|w_{1}|}%
^{2}-{|w_{2}|}^{2},0\right\}  ^{m}\leq 1$ holds for any $\vec w\in
\mathbb{S}^{n_t-1}$. Therefore, $
\Ex{\max\left\{\vec{w}^{H}\vec{A}\vec{w},0\right\}^{m}}  = \lambda_{1}^{m} \Ex{\max\left\{  {|w_{1}|}%
^{2}-{|w_{2}|}^{2},0\right\}  ^{m}} \leq \lambda_{1}^{m}
$
together with \eqref{eq:lambda} proves the upper bound.
\end{IEEEproof}
Now we are ready to prove Theorem \ref{theo:scalingIA}.
\begin{IEEEproof}
Consider an arbitrary but fixed user $m\in\mathcal{S}_{b}$ where
$|\mathcal{S}_{b}|$ is non-random and fulfills the feasibility condition
\eqref{eq:maxUE}, for all $b=1,...,K$, with equality. Define an IA solution $\vec{\pi}_{\text{IA}}$
as $|\langle\vec{v}_{m,b},\vec{\pi}_{\text{IA}}(k)\rangle|^{2}=0$, for all
$b = 1,\ldots,K$ and $k\in \set S \setminus \{m\}$. % , such that $\vec{\pi}_{\text{IA}}(m)$ is independent of
 % the effective channels $\vec{h}_{m,b}$ and isotropically distributed. 
For sufficiently high SNR, $R(\vec \pi_{\text{IA}},S,V)$ achieves the
optimal capacity scaling \eqref{eq:partCSI_sched}. Thus, we can use Lemma
\ref{lem:motivation} and get
\begin{equation}
\mathbb{E}\left[  {\Delta r_{m}(\vec{\pi}_{H},\vec{\pi}_{V})}\right]
\geq\mathbb{E}\left[  {\max\left\{  r_{m}(\vec{\pi}_{\text{IA}},\mathcal{S}%
;H)-r_{m}(\vec{\pi}_{\text{IA}},\mathcal{S};V),0\right\}  }\right],
\end{equation}
with
\begin{align*}
r_{m}(\vec{\pi}_{\text{IA}},\mathcal{S};H) &  =\log\left(  1+\frac{\frac
{P\mu_{m,b}^{2}}{|\mathcal{S}_{b}|}|\langle\vec{h}_{m,b},\vec{\pi}_{\text{IA}%
}(m)\rangle|^{2}}{1+\sum_{l=1}^{K}\sum_{\multi{k\in\mathcal{S}_{l}}{k\neq m}%
}\frac{P\mu_{m,l}^{2}}{|\mathcal{S}_{l}|}|\langle\vec{h}_{m,l},\vec{\pi
}_{\text{IA}}(k)\rangle|^{2}}\right)  \\
r_{m}(\vec{\pi}_{\text{IA}},\mathcal{S};V) &  =\log\left(  1+\frac
{P}{|\mathcal{S}_{b}|}\mu_{m,b}^{2}|\langle\vec{v}_{m,b},\vec{\pi}_{\text{IA}%
}(m)\rangle|^{2}\right).
\end{align*}
Similar to \eqref{eq:effH} and  \eqref{eq:effV} we define the following variables (index $m$ omitted)
\begin{align*}
\Phi_{b,k}^{\ast} &  =\frac{P\mu_{m,b}^{2}}{|\mathcal{S}_{b}|}|\langle\vec
{h}_{m,b},\vec{\pi}_{\text{IA}}(k)\rangle|^{2}\\
\Psi_{b,k}^{\ast} &  =\frac{P\mu_{m,b}^{2}}{|\mathcal{S}_{b}|}|\langle\vec
{v}_{m,b},\vec{\pi}_{\text{IA}}(k)\rangle|^{2}\\
\Delta_{b,k}^{\ast} &  =\frac{|\mathcal{S}_{b}|}{P\mu_{m,b}^{2}}\max\left\{
\Phi_{b,k}^{\ast}-\Psi_{b,k}^{\ast},0\right\}, 
\end{align*}
which can be interpreted as the effective receive SNR of user $m$
for the IA solution. Using this notation the rate gap can be written
in compact form, 
\begin{equation}
  \Delta r_{m}(\vec{\pi}_{H},\vec{\pi}_{V})
  \geq\max\left\{  \log\left(  1+\sum_{l=1}^{K}\sum_{k\in\mathcal{S}_{l}}
\Phi_{l,k}^{\ast}\right)  -\log\left(  1+\sum_{l=1}^{K}\sum_{\multi{k\in\mathcal{S}_{l}}{k\neq m}
}\Phi_{l,k}^{\ast}\right)  -\log\left(  1+\Psi_{b,m}^{\ast}\right)
,0\right\}
\end{equation}
and since $\max\{a-b-c,0\}\geq\max\{a-b,0\}-c$ for $c>0$, the rate gap for
user $m$ is bounded from below by
\begin{align}
\Delta r_{m}(\vec{\pi}_{H},\vec{\pi}_{V})  &  \geq\log\left(  1+\frac{\max\left\{  \Phi_{l,m}^{\ast}-\Psi_{b,m}%
^{\ast},0\right\}  }{1+\Psi_{b,m}^{\ast}}\right)  -\log\left(  1+\sum
_{l=1}^{K}\sum_{\multi{k\in\mathcal{S}_{l}}{k\neq m}
}\Phi_{l,k}^{\ast}\right)  \nonumber\\
&  \geq\underbrace{\log\Biggl(  1+\frac{P\mu_{m,b}^{2}}{|\mathcal{S}_{b}|\left(
1+P\mu_{m,b}^{2}\right)  }\Delta_{b,m}^{\ast}\Biggr)}_{:= \set A}  - \underbrace{\log\left(
1+\sum_{l=1}^{K}\sum_{\multi{k\in\mathcal{S}_{l}}{k\neq m}%
}\Phi_{l,k}^{\ast}\right)}_{:=\set B}  . \label{eq:AB}
\end{align}
To bound $\Ex{\Delta r_{m}(\vec{\pi}_{H},\vec{\pi}_{V})}$ from below,
we will derive a lower bound on the expected value of $\set A$ and an
upper bound on the expected value of $\set B$.

We start with the upper bound for  $\set B$. Since, $\vec \pi_{IA}$ is an IA solution according to $V$, we have $|\langle\vec{h}_{m,b},\vec{\pi}_{\text{IA}}(k)\rangle|^{2}\leq
Z$, for $k\neq m$, where $Z=\underset{\vec{v}\in\mathcal{V}}{\min}(1-|\langle\vec{h}
_{k,l},\vec{v}\rangle|^{2})$ is the quantization error (defined in Lemma
\ref{lem:q_error}) under RVQ. By Lemma \ref{lem:q_error} and the 
Cauchy-Schwarz inequality we have
\begin{equation}
\mathbb{E}\left[  \Phi_{l,k}^{\ast}\right]  \leq\frac{P}{|\mathcal{S}_{l}%
|}\sqrt{\mathbb{E}\left[  \mu_{m,b}^{4}\right]  }\sqrt{\mathbb{E}\left[
Z^{2}\right]  } \leq \frac{P}{|\mathcal{S}_{l}|}\sqrt{\mathbb{E}\left[  \mu_{m,b}^{4}\right]  }2^{\frac{-B}{n_{t}-1}},\;\forall k\neq m ,l. 
\end{equation}
Using Jensen's inequality and $\frac{|\set S_l|-1}{|\set S_l|} \leq
1$, for all $l$, we obtain the upper bound 
\begin{equation}
  \label{eq:boundB}
  \Ex{\set B} \leq \log\left(1 + KP\sqrt{\mathbb{E}\left[  \mu_{m,b}^{4}\right]  }2^{\frac{-B}{n_{t}-1}}\right).
\end{equation}
To lower bound $\set A$ we define the positive random variable
\begin{align}
  Y &  :=\left(  \Delta_{b,m}^{\ast}\right)  ^{2}  =\max\left\{ | \langle\vec{h}_{m,b},\vec{\pi}_{\text{IA}}(k)\rangle
|^{2}-|\langle\vec{v}_{m,b},\vec{\pi}_{\text{IA}}(k)\rangle|^{2} , 0 \right\}^{2}, 
\end{align}
where the mapping between $Y$ and $\Delta_{b,m}^{\ast}$ is bijective, since
$\Delta_{b,m}^{\ast}$ is positive per definition. Taking expectation conditioned on
$\mu_{m,b}$ and $\vec{h}_{m,b}$ (denoted $\mathbb{E}\left[  \cdot|\mu_{m,b},\vec
{h}_{m,b}\right]  :=\mathbb{E}_{|\mu,\vec{h}}\left[  \cdot\right]  $) and
using Lemma \ref{lem:lowerBound} with the concave function $f( x)=\log\left(  1+\sqrt{x}\right)$ we get
\begin{align*}
\mathbb{E}_{|\mu,\vec{h}}\left[ \set A \right] & =  \mathbb{E}_{|\mu,\vec{h}}\left[  \log\left(  1+\frac{P\mu_{m,b}^{2}%
}{|\mathcal{S}_{b}|\left(  1+P\mu_{m,b}^{2}\right)  }\Delta_{b,m}^{\ast
}\right)  \right]  \\
&  = \mathbb{E}_{|\mu,\vec{h}}\left[  \log\left(  1+\frac{P\mu_{m,b}^{2}%
}{|\mathcal{S}_{b}|\left(  1+P\mu_{m,b}^{2}\right)  }\sqrt{Y}\right)  \right]
\\
&  \geq \frac{1}{c_{1}}\left(  1-\sqrt{\frac{\mathbb{E}_{|\mu,\vec{h}}\left[
Y^{2}\right]  }{c_{1}\mathbb{E}_{|\mu,\vec{h}}\left[  Y\right]  ^{2}}}\right)
\log\left(  1+c_{1}\frac{P\mu_{m,b}^{2}}{|\mathcal{S}_{b}|\left(  1+P\mu
_{m,b}^{2}\right)  }\sqrt{\mathbb{E}_{|\mu,\vec{h}}\left(  Y\right)  }\right).
\end{align*}
It remains to compute the first and second moment of $Y$. 
Since, conditioned on $\mu_{m,b} $ and $\vec{h}_{m,b}$ the beamformer $\vec{\pi
}_{\text{IA}}(m)$ is isotropic distributed, we have by Lemma
\ref{lem:ExChor2} (first step, $n=2$) and Lemma \ref{lem:q_error} (last step)
\begin{align*}
\mathbb{E}_{|\mu,\vec{h}}\left(  Y\right)   &  =\mathbb{E}_{|\mu,\vec{h}%
}\left[  \max\left\{  |\langle\vec{h}_{m,b},\vec{\pi}_{\text{IA}}%
(m)\rangle|^{2}-|\langle\vec{v}_{m,b},\vec{\pi}_{\text{IA}}(m)\rangle
|^{2},0\right\}  ^{2}\right]  \\
&  \geq\frac{4\cdot2^{-n_{t}}}{n_{t}\left(  n_{t}+1\right)  }\mathbb{E}%
_{|\mu,\vec{h}}\left[  \underset{\vec{v}\in\mathcal{V}}{\min}(1-|\langle
\vec{h}_{m,b},\vec{v}\rangle|^{2})\right]  \\
&  =\frac{4\cdot2^{-n_{t}}}{n_{t}\left(  n_{t}+1\right)  }\mathbb{E}%
_{|\mu,\vec{h}}\left(  Z\right)  \\
&  \geq\frac{4\cdot2^{-n_{t}}}{n_{t}\left(  n_{t}+1\right)  }\frac{n_{t}%
-1}{n_{t}}2^{\frac{-B}{n_{t}-1}}.
\end{align*}
Again by Lemma \ref{lem:ExChor2} (first step) and Lemma
\ref{lem:q_error} (second step) we have 
\begin{align}
\mathbb{E}_{|\mu,\vec{h}}\left(  Y^{2}\right)   &  \leq\mathbb{E}_{|\mu
,\vec{h}}\left[  Z^{2}\right]  
  \leq\left(  \frac{n_{t}}{n_{t}-1}\mathbb{E}_{|\mu,\vec{h}}\left[
{Z}\right]  \right)  ^{2}. 
\end{align}
Such that, 
\begin{multline}
 \mathbb{E}_{|\mu,\vec{h}}\left[ \set A\right] 
 % & \geq \frac{1}{c_{1}}\left(  1-\sqrt{\frac{\mathbb{E}_{|\mu,\vec{h}}\left[
% Y^{2}\right]  }{c_{1}\mathbb{E}_{|\mu,\vec{h}}\left[  Y\right]  ^{2}}}\right)
% \log\left(  1+\frac{c_{1}P\mu_{m,b}^{2}}{|\mathcal{S}_{b}|\left(  1+P\mu
% _{m,b}^{2}\right)  }\sqrt{\mathbb{E}_{|\mu,\vec{h}}\left(  Y\right)  }\right)
% \\
%&   \geq\frac{1}{c_{1}}\left(  1-\frac{n_{t}}{\sqrt{c_{1}}\left(
% n_{t}-1\right)  \frac{4\cdot2^{-n_{t}}}{n_{t}\left(  n_{t}+1\right)  }%
% }\right)  \log\left(  1+\frac{c_{1}P\mu_{m,b}^{2}}{|\mathcal{S}_{b}|\left(
% 1+P\mu_{m,b}^{2}\right)  }\sqrt{\frac{4\cdot2^{-n_{t}}}{n_{t}\left(
% n_{t}+1\right)  }\frac{n_{t}-1}{n_{t}}}2^{\frac{-B}{2\left(  n_{t}-1\right)
% }}\right)  \\
  \geq \frac{1}{c_{1}}\left(  1-\frac{n_{t}^{2}\left(  n_{t}+1\right)  }%
{4\sqrt{c_{1}}\left(  n_{t}-1\right)  2^{-n_{t}}}\right)  \\ \log\left(
1+\frac{c_{1}P\mu_{m,b}^{2}}{|\mathcal{S}_{b}|\left(  1+P\mu_{m,b}^{2}\right)
}\sqrt{\frac{4\cdot2^{-n_{t}}(n_{t}-1)}{n_{t}^{2}\left(  n_{t}+1\right)  }%
}2^{\frac{-B}{2\left(  n_{t}-1\right)  }}\right). \label{eq:boundA}
\end{multline}
Plugging \eqref{eq:boundA} and \eqref{eq:boundB} in \eqref{eq:AB} and
taking expectation with respect to $\mu_{m,b}$ the
claim follows. 
\end{IEEEproof}

\end{document}